\documentclass{article}
\usepackage{arxiv}

\usepackage{setspace}
\usepackage{hyperref}
\usepackage[table]{xcolor}
\usepackage{atbeginend}
\usepackage{float}
\usepackage{pdfpages}
\usepackage{longtable}
\usepackage{xspace}
\usepackage{url}
\usepackage{showulab}
\usepackage{epsfig}
\usepackage{multirow}
\usepackage{caption}
\usepackage{subcaption}
\usepackage{graphicx}
\graphicspath{{./Graphs/}}
\usepackage{tabularx}
\usepackage{placeins} %used to keep floats in same section of paper with command \FloatBarrier

\usepackage{doi}
\usepackage{booktabs}

\newcommand{\auc}{AUC\xspace}
\newcommand{\aucs}{AUCs\xspace} 

\newcommand{\VV}{V\&V\xspace} %xspace adds or remove space after word as necessary
\newcommand{\M}{MBSwE\xspace}

\newcommand{\MBSE}{MBSwE\xspace}
\hypersetup{
pdftitle={On the Practices of Autonomous Systems Development: Survey-based Empirical Findings},
pdfsubject={q-XXXX.NC, q-XXXX.QM},
pdfauthor={Katerina Goseva-Popstojanova,Denny Hood,Johann Schumann,Noble Nkwocha}
pdfkeywords={Autonomous systems, Development practices, Reuse, Processes and standards, Verification and validation, software bugs},
}

\begin{document}

\title{On the Practices of Autonomous Systems Development: Survey-based Empirical Findings}
\date{}

\author{ 
        {
\hspace{1mm}Katerina Goseva-Popstojanova}\\
	Department of Computer Science \& Electrical Engineering\\
	West Virginia University, Morgantown, WV 26506, USA \\
	\texttt{Katerina.Goseva@mail.wvu.edu} \\
	\And
	{Denny Hood} \\
	Department of Computer Science \& Electrical Engineering\\
	West Virginia University, Morgantown, WV 26506, USA \\
	\texttt{dlh0038@mix.wvu.edu} \\
	\AND
	{Johann Schumann} \\
	KBR/Wyle LLC, NASA Ames Research Center, Moffett Field, CA 94035, USA\\
	\texttt{johann.m.schumann@nasa.gov} \\
	\And
	{Noble Nkwocha} \\
	NASA Independent Verification \& Validation Facility, Fairmont, WV 26554, USA \\
	\texttt{noble.n.nkwocha@nasa.gov}
}

\maketitle

\begin{abstract}
Autonomous systems have gained an important role in many industry domains
and are beginning to change everyday life.
However, due to dynamically emerging applications and often proprietary 
constraints, there is a lack of information about the practice 
of developing autonomous systems. This paper presents the first part of the longitudinal study focused on establishing state-of-the-practice,
identifying and quantifying the challenges and benefits,
identifying the processes and standards used, 
and exploring verification and validation (V\&V) practices used for 
the development of autonomous systems.
The results presented in this paper are based on data about software systems that have autonomous functionality  and may employ model-based software engineering (MBSwE) and reuse. 
These data were collected using an anonymous online survey that was administered in 2019 and were provided by 
experts with experience in development of autonomous systems and / or the use of MBSwE.   
Our current work is focused on repeating the survey to collect more recent data and discover how the development of autonomous systems has evolved over time.
\end{abstract}

%=====================================================================
\section{Introduction}
\label{ch:introduction}

The advancements of computing technologies (i.e., sensors, embedded processing, computer vision, hardware acceleration, machine learning) and communication technologies have enabled and motivated many organizations to work on development of autonomous systems. 
Domains such as space exploration, military, agriculture, and healthcare already have numerous use cases for autonomous systems. Many automotive companies, such as ArgoAI, Audi, Baidu, Cruise, Mercedes-Benz, Nissan, Tesla, Uber, and Waymo, have embraced and made enormous investments in self-driving cars technology \cite{liu2020computing-State-of-the-art-Challenges}.
However, extremely high levels of complexity and safety criticality present fundamental challenges \cite{harel2021creating}.
Autonomous systems have complex interactions -- they perceive the environment and execute tasks without detailed programming or under direct human control. Unlike automated systems, which execute a carefully engineered sequence of actions that cannot be changed, autonomous systems self-govern their course of action, that is, they are meant to understand and decide how to execute tasks based on goals, skills, and learning experience \cite{ebert2019validation}.

NASA defines autonomy as the ability of a system to achieve goals while operating independently of external control \cite{hall2015roadmap}. 
An autonomous system can, and in the case of space domain often must, make decisions and take actions with little or no human involvement \cite{cardoso2021review-space}. 
Autonomous systems range from model-based diagnostics and prognostics to machine learning (ML) and artificial intelligence (AI) based systems, and may be both safety and mission critical, where failures can place human life and the mission in jeopardy. 
For example, in spite of the advances in the domain of autonomous vehicles, catastrophic accidents have happened, such as an autonomous car misinterpreting a white truck as a white cloud, and another one overlooking pedestrians on a road, which led to a fatal accident \cite{ebert2019validation}. More recently, Tesla has recalled 362,758 vehicles and warned that the driver assistance software, marketed as ``Full Self-Driving Beta'', may cause crashes \cite{TeslaRecall-2023}.

Autonomous systems are composed of Autonomous Components (AUCs) that provide autonomous capabilities and support autonomous operations. Some examples of AUCs are vision-based navigation, system health management and prognostics, flight management systems that can operate without human intervention, and subsystems of deep space missions which perform complex operations without direct remote operator control.
AUCs can implement different autonomous tasks related to (1) information acquisition, (2) information analysis, (3) decision and action selection, and (4) action implementation \cite{parasuraman2000model}.

%%%%%%%%%%%%%%%%%%%%%%%%%%%%%%%%%%%%%%%%%%%%%%%%%%%
% Verification and validation 
%%%%%%%%%%%%%%%%%%%%%%%%%%%%%%%%%%%%%%%%%%%%%%%%%%%
Autonomous systems impose novel and hard challenges for Verification and Validation (V\&V). 
For example, based upon their literature review of verification and validation techniques for space autonomous systems \cite{cardoso2021review-space}, the authors concluded that verification and validation are still technological challenges and emphasized the need for providing assurances and guarantees towards reliable missions.
Another literature review work focusing on testing autonomous vehicles \cite{karunakaran2022challengesTesting} has identified four categories of significant challenges: 
complex and unpredictable environment, 
concerns with existing automotive testing methods, 
incompatible safety standards and certification, and 
machine learning induced challenges.
To achieve dependable and trustworthy autonomous systems, intelligent V\&V techniques that cover dynamic changes and learning are needed \cite{ebert2019validation}. 

Software standards for safety-critical systems, like DO-178C \cite{DO-178C},
DO-331 \cite{DO331},
 or ISO 26262 \cite{ISO.26262} for the automotive industry have been 
established at the time when our survey was held in 2019.
These standards aim at complex, safety-critical systems that may have
some automation capabilities, promote concpets like
Reusable Software Components (RSCs) \cite{DO-178C}, and focus on model-based
development and verification like the DO-331 supplement to DO-178C \cite{DO331}.
Guidelines and standards for development and \VV of autonomous systems, e.g., 
\cite{EASAL1concept2021,EASAL1-2concept2024,IEEE7009_2024}, have been published several years after our
survey has been administered.

Due to the emerging nature of autonomous systems and their proprietary nature, there is a lack of information about the state-of-the-practice, and the actual benefits and challenges observed in practice. 
{This paper presents the first part of the longitudinal study focused on addressing this gap. The results are based on data collected from April 25, 2019 to June 20, 2019 by means of an anonymous survey.}
The survey focused on autonomous systems from different industry domains worldwide, and included aspects related to Model-based Software Engineering (MBSwE) and reuse. Out of 129 respondents to the survey, 110 used autonomous systems and/or MBSwE in their projects and answered questions beyond the first 
survey question. To assure that respondents' answers are not biased, the survey was administered anonymously. 
The survey had six specific sections addressing multiple research questions, which are listed in Table~\ref{table:research_questions}.
%%%%%%%%%%%%%%%%%%%%%%%%%%%%%%%%%%%%%%%%%%%
% Contributions
%%%%%%%%%%%%%%%%%%%%%%%%%%%%%%%%%%%%%%%%%%%
{The detailed analysis of the corresponding results led to the following contributions, pertinent to the time period in which the data were collected (i.e., 2019):} 

\begin{description}
\item[C1.] established the {\bf state-of-the-practice of developing autonomous systems} (i.e., answered the ``Where'', ``What'', ``How'', ``Why'', and ``Who'' questions) based on expert opinions collected using an anonymous survey, 
\item[C2.] identified and quantified the {\bf challenges and benefits of autonomy and reuse},
\item[C3.] identified the {\bf processes and standards} used to develop autonomous systems, and
\item[C4.] explored the {\bf verification and validation used for the autonomy, models, and reuse}.
\end{description}

%------------------------------------------------------------------
\begin{table}[ht]
\centering
\caption{Survey Sections with corresponding research questions and contributions}
\label{table:research_questions}
\begin{tabular}{|p{1cm}|p{13cm}|c|}
\hline
%--------------------------------------------------------------
\multicolumn{2}{|c|}{\cellcolor{lightgray}\textbf{Survey Section 1: Where? What? How? and Who?}} & \cellcolor{lightgray}\  \\ \hline
RQ1a & Which areas of industry are using autonomous systems and/or \MBSE?                                     &\multirow{9}{*}{\centering C1}     \\ \cline{1-2}
RQ1b & What  is  the  level  of  safety  criticality  of  the  applications where autonomous systems and/or MBSwE are being used?& \\ \cline{1-2}
RQ1c & What are the programming languages used during the development and deployment of the project?&                              \\ \cline{1-2}
RQ1d & How  was  the  code for autonomous functionality during development and deployment developed?&                     \\ \cline{1-2}
RQ1e & Was special hardware and/or cloud used for code implementing autonomous functionality?&                                    \\ \cline{1-2}
RQ1f & What were the respondents' roles in the project?                                      &                                      \\ \hline\hline
%--------------------------------------------------------------
\multicolumn{2}{|c|}{\cellcolor{lightgray}\textbf{Survey Section 2: Details on Autonomy (for each AUC in the project)}} & \cellcolor{lightgray}\                                                                                  \\ \hline
RQ2a & Was the AUC developed using \MBSE and which \MBSE tools were used?          &\multirow{5}{*}{\centering C2}     \\ \cline{1-2}
RQ2b & What is the level of autonomy of \auc?                                                                 &     \\ \cline{1-2}
RQ2c & What algorithms and modeling paradigms were used to develop \auc?                                      &     \\ \cline{1-2}
RQ2d & How were the requirements for the \auc specified?                                                      &     \\ \cline{1-2}
RQ2e & What were the challenges associated with the \auc?                                                     &     \\ \hline \hline
%--------------------------------------------------------------
\multicolumn{2}{|c|}{\cellcolor{lightgray}\textbf{Survey Section 3: Details on Reuse of Software Artifacts}}          & \cellcolor{lightgray}\                                                                 \\ \hline
RQ3a & Which artifacts were reused and to what extent?                                                   &\multirow{4}{*}{\centering C3}                          \\ \cline{1-2}
RQ3b & Were there any negative aspects of reuse?                                                                            &      \\ \cline{1-2}
RQ3c & What were the difficulties due to reuse?                                                                             &      \\ \cline{1-2}
RQ3d & What were the benefits of reuse?                                                                                     &      \\ \hline \hline
\multicolumn{2}{|c|}{\cellcolor{lightgray}\textbf{Survey Section 4: Processes and Standards}}                       & \cellcolor{lightgray}\                                                        \\ \hline
RQ4a & Which life-cycle model was used?                                                                          &\multirow{4}{*}{\centering C3}                 \\ \cline{1-2}
RQ4b & Which modeling standards and coding standards were used by the projects?                                              &     \\ \cline{1-2}
RQ4c & Did the system go through a certification process and was the \auc part of the certified system?       &    \\ \hline \hline
\multicolumn{2}{|c|}{\cellcolor{lightgray}\textbf{Survey Section 5: Verification \& Validation}}                            & \cellcolor{lightgray}\                          \\ \hline
RQ5a & Which quality attributes were verified and validated?                                                    &\multirow{5}{*}{\centering C4}                  \\ \cline{1-2}
RQ5b & How were the models verified \& validated?                                                                            &     \\ \cline{1-2}
RQ5c & How were the \aucs verified \& validated during development?                                           &      \\ \cline{1-2}
RQ5d & How was runtime behavior of \aucs monitored/assured?                                                   &     \\ \cline{1-2}
RQ5e & Were the reused artifacts verified \& validated?                                                       &                    \\ \hline \hline
\multicolumn{2}{|c|}{\cellcolor{lightgray}\textbf{Survey Section 6: Bugs}}                                                                          & \cellcolor{lightgray}\                        \\ \hline
RQ6a & Where there any bugs specific to autonomous functionality, model-based approach, and/or reuse?        &\multirow{2}{*}{\centering C4}          \\ \hline
\end{tabular}
\end{table}
%%%%%%%%%%%%%%%%%%%%%%%%%%%%%%%%%%%%%%%%%%% 

The work presented in this paper is comprehensive with respect to autonomous systems in general, from multiple domains, throughout the life cycle, and also incorporates the use of MBSwE and reuse. 
Most of the related studies focused on autonomous systems were based on literature review and systematization.
Some of these studies were focused only on 
one domain (i.e., Unmanned Aerial Vehicles (UAV) \cite{chen2009survey} or
Space \cite{cardoso2021review-space} or 
Autonomous Vehicles (AV) \cite{liu2020computing-State-of-the-art-Challenges,karunakaran2022challengesTesting,tahir2020coverage-safety-AV,gao2021autonomous-Security}),
a particular aspect of the development, like verification and validation \cite{karunakaran2022challengesTesting,tahir2020coverage-safety-AV,schumann2010application,schumann2011verification,schumann2013software,Nikora2018assurance,song2021concepts-PerRuneson},
or particular quality attribute (e.g., safety \cite{tahir2020coverage-safety-AV} or security \cite{gao2021autonomous-Security,jahan2019securityLiteratureReview}).
Different from most of the related works, our findings are based on the practical experiences of experts on autonomous systems,  both from academia and industry. 
Only several prior works were based on experts practical experiences: 
our previous work which was focused only on MBSwE \cite{goseva2016survey},
another work based on a survey \cite{lavallee2006intelligent} published in 2006 which was focused only on space domain, and
a more recent work that addressed only testing of autonomous systems and was based on focus discussions and interviews  with a small sample of experts \cite{song2021concepts-PerRuneson}.

The remainder of the paper is organized as follows: 
Section~\ref{ch:relwork} discusses the related works.
Section~\ref{ch:survey} presents an overview of the survey, its design, and how the survey was administered. 
Detailed analysis of the results of the survey are discussed in Section~\ref{ch:analysis}.
The threats to validity are described in Section~\ref{ch:ThreatsToValidity}.
Finally, Section~\ref{ch:summary} provides a summary of the observations and recommendations.

%--------------------------------------------------------------
%--------------------------------------------------------------
\section{Related Work}
\label{ch:relwork}
Due to its emerging nature, many papers on autonomy have been published. 
We start from the related works that were concerned with the level of autonomy and the types of tasks (i.e., information processing categories) \cite{sheridan1992telerobotics,parasuraman2000model,lavallee2006intelligent,sae2014taxonomy,proud2003methods,chen2009survey}.
Different levels of autonomy were first introduced by Sheridan et al. \cite{sheridan1992telerobotics}. 
Parasuraman et al. \cite{parasuraman2000model} introduced a framework for types and levels of autonomy that act as basis to determine the extent of autonomy for different tasks. The authors introduced four distinct information processing categories (i.e., tasks) of autonomy: information acquisition, information analysis, decision and action selection, and action implementation. The level of autonomy was defined on a scale from 1 to 10, where 1 is full human decision making and 10 is full computer decision making; values in between represented combinations of different levels of human and computer interactions. Different levels of autonomy can be applied to each task. 
LaVallee et al. \cite{lavallee2006intelligent} proposed the following six levels of autonomy tailored to space domain: ``manual'', ``automatic notification'', ``intelligent reasoning on ground with human control'', ``intelligent reasoning on ground with autonomous control'', ``intelligent reasoning onboard'', and ``autonomous thinking spacecraft''.
The Society of Automotive Engineers (SAE) has defined six levels of vehicle autonomy, with Level 0 (L0) being the lowest (i.e., no autonomy) and Level 5 (L5) being the highest (i.e., full autonomy in any driving environment) \cite{sae2014taxonomy}. 
Most of the modern vehicles have Level 1 autonomy, with at least one autonomous feature (e.g., braking, acceleration assistance, or cruise control). Vehicles with Level 2 autonomy use a combination of autonomous features to manage steering, acceleration and braking, but the driver must remain engaged and monitor the environment at all times. GM's Super Cruise and Tesla's Full Self-Driving are examples of Level 2 systems. Mercedes-Benz became the first automaker certified to sell vehicles with SAE Level 3 autonomous technology, which requires a human driver to be ready to take control over the vehicle if necessary. According to Mercedes-Benz, Level 4 autonomy may be possible by the end of the decade \cite{Mercedes-US-News-2023}. 
Proud et al. \cite{proud2003methods} developed the NASA's SMART (Spacecraft Mission Assessment and Re-planning Tool) as a prototype to functionally decompose a flight management system with a suitable level of autonomy for the desired functionality. The authors introduced the Level of Autonomy Assessment Tool which was designed to find the optimum level of autonomy that minimizes the cost, and maximizes safety and efficiency.
To measure the level of autonomy, Chen et al. \cite{chen2009survey} presented concepts related to autonomous systems, such as control level metrics. The architecture of autonomous UAV systems were divided into three levels: execution, coordination, and organization. 

Based on published related works, the work presented in  \cite{liu2020computing-State-of-the-art-Challenges} summarized the state-of-the-art of computing systems for AV. That work considered seven performance metrics (i.e., accuracy, timeliness, power, cost, reliability, privacy, and security) and nine key technologies (i.e., sensors, data source, autonomous driving applications, computation hardware, storage, real-time operating systems, middleware systems, vehicular communication, and security and privacy). In addition, it identified twelve challenges to realizing autonomous driving: artificial intelligence for AVs, multisensors data synchronization, failure detection and diagnostics, how to deal with normal–abnormal, cyberattack protection, vehicle operating system, energy consumption, cost, how to benefit from smart infrastructure, dealing with human drivers, experimental platform, and physical worlds coupling. 

Many related works were focused on different aspects of verification and validation of autonomous systems  \cite{schumann2010application,schumann2011verification,schumann2013software,Nikora2018assurance,song2021concepts-PerRuneson,cardoso2021review-space,karunakaran2022challengesTesting}.
Schumann et al. \cite{schumann2010application} presented a literature review of the use of neural networks in high assurance systems of various fields of study. 
They concluded that traditional verification and validation for safety-critical code is insufficient for neural network applications.
Techniques for analysis and verification and validation of the System Health Management (SHM) were presented in \cite{schumann2011verification}. Specifically, a combination of n-factor combinatorial exploration and Monte Carlo techniques were used, which allowed for detection of potential weaknesses and unwanted parameter sensitivity in the health model.
Schumann et al. \cite{schumann2013software} proposed a Bayesian method to diagnose and avert software faults in real-time using Software Health Management (SWHM). Using a two stage verification and validation process, both the model and code levels of a safety critical component were analyzed. The authors concluded that SWHM can provide an additional layer of safety during runtime, but was not suitable for replacement of verification and validation, and certification of the system.
A paper by Nikora et al. \cite{Nikora2018assurance} investigated the verification and validation techniques for systems with autonomous capabilities in the following areas: diagnostic model, diagnostic engine, and combination of model and engine. The authors noted that test procedures and expected results about a component's nominal functionality were more easily obtained than the off-nominal behavior.
A recent review of verification and validation techniques for space autonomous systems discussed the model checking, theorem proving, runtime verification, software testing, and verification and validation of machine learning \cite{cardoso2021review-space}. The authors concluded that verification and validation were still challenges for space autonomous systems. 
Another recent literature review by Karunakaran et al. \cite{karunakaran2022challengesTesting}, which was focused on testing AV, identified four categories of significant challenges: complex and unpredictable world, concerns with existing automotive testing methods, incompatible safety standards and certification, and machine learning induced challenges. 
The work by Song et al. \cite{song2021concepts-PerRuneson} was also focused on testing of autonomous systems, and in addition to literature review used focus group discussions consisting of 8 participants and semi-structured interviews with 5 participants. That work classified the challenges of testing autonomous systems to four categories: unpredictable environment; system and scenario complexity; data accessibility; and missing standards and guidelines. It also classified the available techniques, approaches, and practices for testing of autonomous systems. 

%%%%%%%%%%%%%%%%%%%%%%%%%%%%%%%%%%%%%%%%%%%%%%%%%%%%%%%%%%%
%             Empirical studies of software bugs          %
%%%%%%%%%%%%%%%%%%%%%%%%%%%%%%%%%%%%%%%%%%%%%%%%%%%%%%%%%%%

Even though the empirical analysis of software bugs is a very active
research area, only several recent papers were focused on studying  software bugs in systems with autonomous functionality \cite{garcia2020-AV,tang2021-AV,taylor2021-UAS,wang2021-UAS,goseva2023COMPSAC}. 
Of these, two works were focused on AV bugs \cite{garcia2020-AV,tang2021-AV}. 
Garcia et al. \cite{garcia2020-AV} presented an empirical study of 499 bugs of the open source autonomous driving systems Apollo and Autowarein, and using manual analysis classified the root causes and symptoms of the bugs, as well as identified the autonomous components affected by these bugs. 
The study by Tang et al. \cite{tang2021-AV} was based on the open-source driver assistant system OpenPilot, for which the authors collected 235 bugs and classified them into five categories.  
Another two papers studied the bugs in UAS \cite{taylor2021-UAS,wang2021-UAS}, both based on two open source software suites PX4 (capable of controlling drones) and ArduPilot (capable of controlling unmanned vehicle systems such as drones, aircrafts, helicopters, ground rovers, boats, submarines, and antenna trackers).
Taylor et al. \cite{taylor2021-UAS} investigated 277 firmware bugs in the ArduPilot and PX4 code bases, with a focus on the root causes, symptoms, reproducability, and location of the bugs. 
Wang et al. \cite{wang2021-UAS} extracted 569 bugs of ArduPilot and PX4, and using a manual labeling process identified 168 UAS-specific bugs whose root causes were classified into 8 categories.
%%%%%%%%%%%%%%%%%%%%%%%%%%%%%%%%%%%%%%%%%%%%%%%%
In our work \cite{goseva2023COMPSAC}, we investigated software changes and bugs in the Autonomy Operating System (AOS) for UAS, which has 26 components with a total of 772 bugs. The results showed that autonomous components were significantly more prone to changes (measured in number of commits and code churn) and fault prone (measured in bugfixes per KLoC) than non-autonomous components. Furthermore, the distribution of the locations of bugs was skewed, both at component and file level (i.e., a small number of components / files contained the majority of bugs).

%%%%%%%%%%%%%%%%%%%%%%%%%%%%%%%%%%%%%%%%%%
%%%%%%      Safety papers         %%%%%%
%%%%%%%%%%%%%%%%%%%%%%%%%%%%%%%%%%%%%%%%%%
While safety considerations in complex hardware-software systems have been
studied for many years, major work on safety for autonomous systems started
after our survey, e.g., \cite{BURTON202010,CA2018,W2020}.

%%%%%%%%%%%%%%%%%%%%%%%%%%%%%%%%%%%%%%%%%%
%%%%%%      Secuirty papers         %%%%%%
%%%%%%%%%%%%%%%%%%%%%%%%%%%%%%%%%%%%%%%%%%

Some works were focused on security aspects of autonomous systems \cite{jahan2019securityLiteratureReview,gao2021autonomous-Security}. 
For example, the literature review paper by Jahan et al. \cite{jahan2019securityLiteratureReview} presented attack models that have been proposed over the years, proposed a taxonomy of attacks on autonomous systems, and identified the research gap that needs to be addressed.
Another work, which was focused specifically on security of autonomous driving \cite{gao2021autonomous-Security}, elaborated the security issues along four dimensions: sensors, operating system, control system, and vehicle-to-everything (V2X) communication.

%%%%%%%%%%%%%%%%%%%%%%%%%%%%%%%%%%%%%%%%%%%%%%%%%%%%%%%%%%%%%%%%%
%%%%%%                      SURVEY                          %%%%%
%%%%%%%%%%%%%%%%%%%%%%%%%%%%%%%%%%%%%%%%%%%%%%%%%%%%%%%%%%%%%%%%%
We next discuss the empirical works that used survey as an instrument to collect the relevant data \cite{goseva2016survey,lavallee2006intelligent}.
These include our prior work which was based on conducting a survey focused only on the use of \MBSE and auto-generated code (AGC) in various industry domains worldwide \cite{goseva2016survey}. Specifically, based on the answers provided by 114 respondents to the survey, we explored the state-of-the-practice, the benefits and challenges, and software assurance of models and AGC. 
The only previous survey related to autonomous systems was conducted by Lavallee et al. \cite{lavallee2006intelligent}. That work was focused only on space domain and presented a categorization of the level of autonomy and the complexity of the implementation, from a single component to an entire flight and ground systems. 
That work used six levels of autonomy and,
based on 88 survey responses for 62 implementations, reported that lower levels of autonomy dominated (45\% of implementations had Level 2 autonomy, while only 3\% had Level 5 and 8\% had Level 6 autonomy). 
%

%%%%%%%%%%%%%%%%%%%%%%%%%%%%%%%%%%%%%%%%%%%%%%%%%%%%%%%%%%%%%%%%%
%%%%%%                      THIS PAPER                      %%%%%
%%%%%%%%%%%%%%%%%%%%%%%%%%%%%%%%%%%%%%%%%%%%%%%%%%%%%%%%%%%%%%%%%
In this paper, we present the results of the anonymous online survey which was used to collect information about software systems that have autonomous functionality, may have employed model-based software engineering and some level of software reuse, for different domains, worldwide. 
{The survey was conducted from April 25, 2019 to June 20, 2019.}
We aimed to fill the existing gaps in the knowledge related to autonomous systems development at that time by  
(C1) assessing the state-of-the-practice using autonomous systems, 
(C2) identifying and quantifying the benefits and challenges of autonomy and reuse, 
(C3) exploring the processes and standards used to develop autonomous systems, and 
(C4) investigating the verification and validation of the models, autonomy, and reuse.
Unlike most of the related work papers which were based on literature review, the findings presented in this paper are based on the practical experience of experts on autonomous systems, both from academia and industry. 

%--------------------------------------------------------------
\section{Survey Design and Execution}
\label{ch:survey}

The survey was developed following the steps outlined in \cite{Kitchenham2008SurveyPaper}: (1) define the goals, (2) transform the goals into research questions, (3) design the questionnaire, (4) evaluate the questionnaire using pilot executions,  (5) execute the survey, and (6) analyze and package results.
Our Survey consisted of 48 multiple choice and free response questions divided into the following sections (see Table~\ref{table:research_questions}):
\begin{enumerate}
\item \textit{``Where'', `What'', ``How'', and ``Who''} section addressed questions about areas of industry, level of safety criticality, programming languages used, how was the code developed and if the special hardware was used, as well as the respondents role in the project.
\item \textit{Autonomy Details} section contained questions about the level of autonomy, modeling paradigms, requirement specification, and challenges developing autonomous systems.
\item \textit{Reuse} section was concerned with the extent of reuse for \auc and non-\auc and the benefits and challenges related to reuse.
\item \textit{Processes and Standards} section focused on lifecycle, modeling, and coding standards, and certification.
\item \textit{Verification and Validation} section concentrated on questions about \VV of the models, autonomy, and reuse.
\item \textit{Bugs} section dealt with bugs related to the models, autonomous functionality, and reuse of software artifacts.
\end{enumerate}

\begin{figure}[th]
\begin{center}
\includegraphics[scale=0.6]{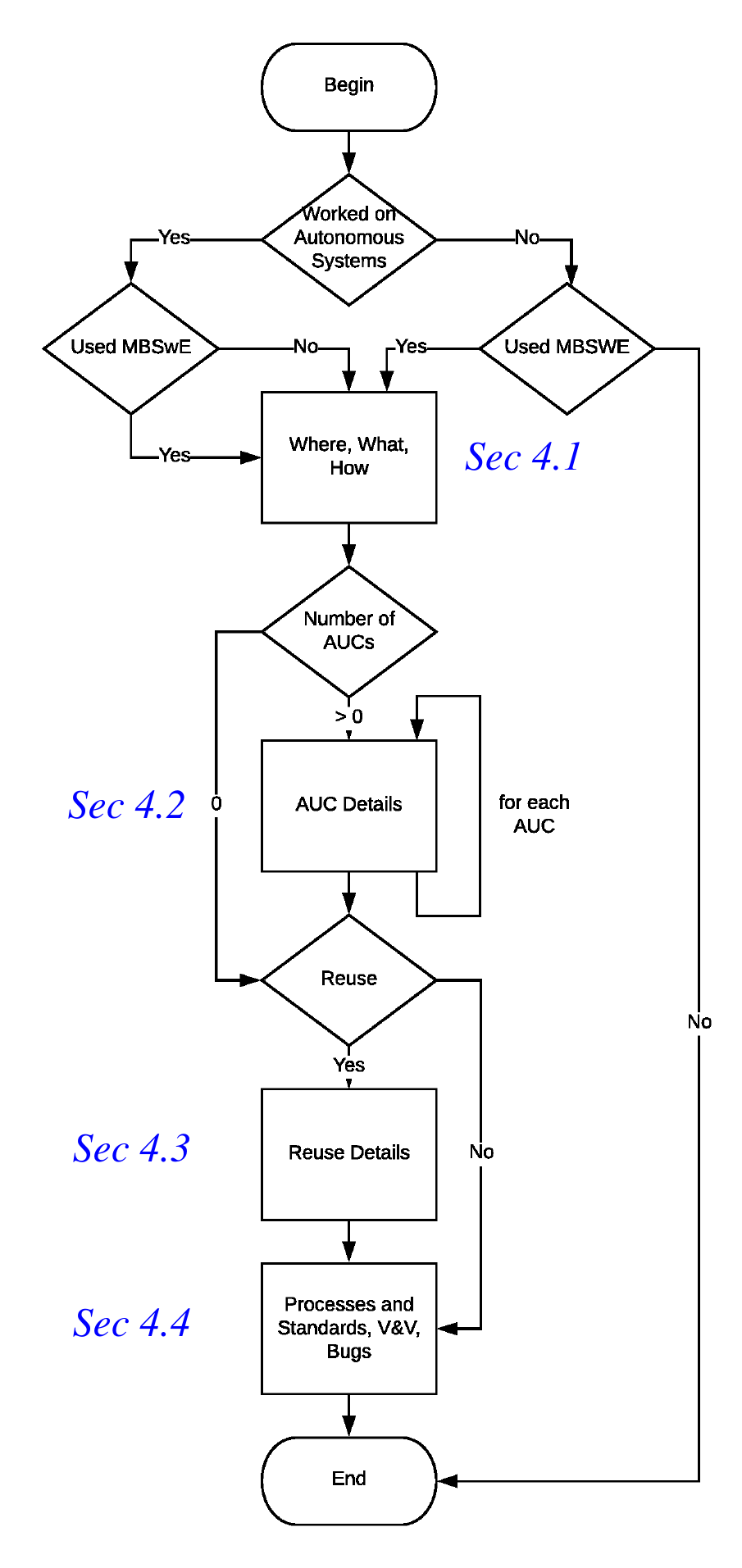}
\end{center}
\caption{Flowchart of our survey}
\label{fig:flowchart}
\end{figure}

The survey was divided into separate pages based on these  sections. Depending on answers to specific questions, respondents were presented with some questions and skipped other questions.  
As can be seen in the flowchart of the survey shown in Figure \ref{fig:flowchart}, the first question was used to determine if the respondent has worked on autonomous systems, followed by the second question to determine if the respondent has used \MBSE. If the respondent neither worked on autonomous systems nor used \MBSE, they were led to the end of the survey. Otherwise, the respondent was asked the first set of questions about ``Where'',  ``What'',  ``How'' of their work. 
Then, the respondents were asked about the number of \aucs they worked on. Based on the response, the questions from the Autonomy Details section were repeated for each \auc. 
Next, the respondents were asked about the existence of reuse in their project. If the respondent chose yes, they were directed to the questions in the Reuse Details section. Otherwise, these questions were skipped. All respondents who were allowed to enter the survey went through the questions from the Processes and Standards, Verification and Validation, and Bugs sections.

The online questionnaire was created using Survey Monkey \cite{surveymonkey}.
A pilot study was used to evaluate the first version of the survey questionnaire. Our colleagues and contacts who had practical experience in developing autonomous systems participated in the pilot study and their comments and suggestions were used to revise and improve the survey questionnaire. 

Invitations to complete the survey were distributed by email to people who work in related fields. The intended respondents were software engineering practitioners and researchers with experience in developing autonomous systems and / or using MBSwE in industry. 
We tried to reach as many potential respondents as possible, using non-probabilistic convenience sampling and snowballing. Techniques included sending invitation messages to academic and industrial contacts of the research team and to relevant mailing lists, and placing advertisements at related conferences and online forums. Additionally, over 300 authors of research papers related to autonomy were contacted via email invitations.
The responses to the survey were collected from April 25, 2019 to June 20,
2019.

Of the 129 respondents who started the survey, 110 respondents have worked on autonomous systems and/or used \M in their projects. These respondents were allowed to enter the survey, that is, answered more than just the first two survey questions. Note that it was not mandatory to answer each question. 

\section{Detailed Analysis of Survey Responses}
\label{ch:analysis}
%=======================================================================

This section presents the analysis of the survey responses provided by the 110 respondents who have used \auc and/or \M in their projects. Since the survey did not require the respondents to answer all questions, for each figure and table, we provide the actual number of respondents and/or the percentages of the total number of respondents answering that specific question.

%%%%%%%%%%%%%%%%%%%%%%%%%%%%%%%%%%%%%%%%%%%%%%%%%%%%%%%%%%%%%%%%%%%%%%%%%%%%%%%%%%%%%%
\subsection{Survey Section 1: Where? What? How? and Who?}
%%%%%%%%%%%%%%%%%%%%%%%%%%%%%%%%%%%%%%%%%%%%%%%%%%%%%%%%%%%%%%%%%%%%%%%%%%%%%%%%%%%%%%
In the first section of the survey, we examined the state-of-the-practice of developing autonomous systems in order to provide answers to ``Where'', ``What'', ``How'', and ``Who'' questions.  

\subsubsection{RQ1a: Which areas of industry are using autonomous systems and/or \MBSE?}

\begin{figure}[ht]
\begin{center}
\includegraphics[trim=0 270 0 230,clip,width=0.75\textwidth]{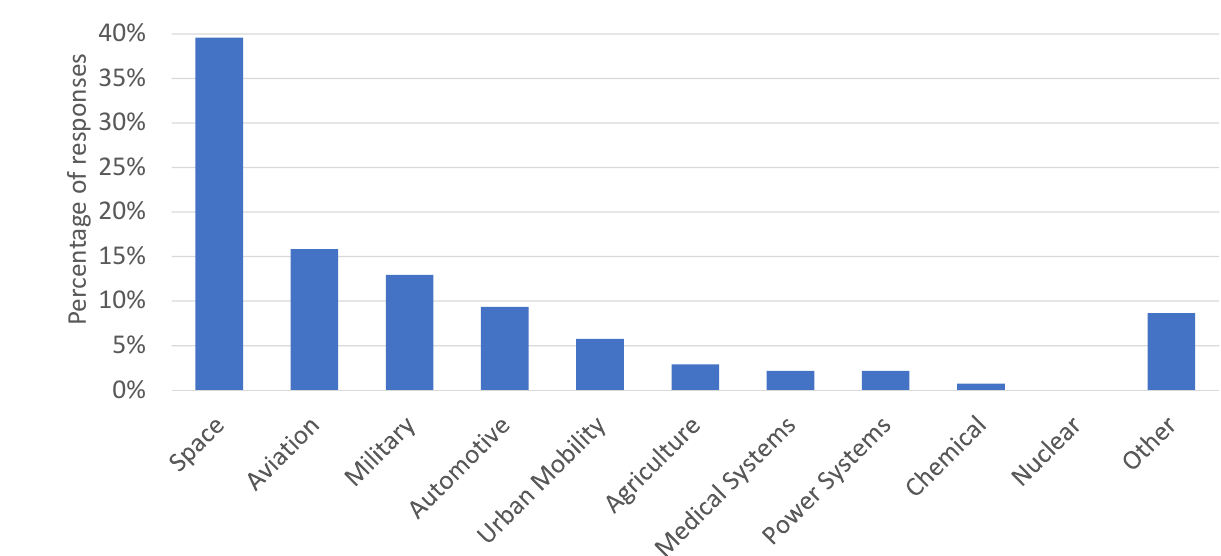}
\end{center}
\caption[Areas of industry that use MBSwE]{Areas of industry to which the respondents to our survey belong. (99 respondents, multiple answers possible)}
\label{fig:targetdomains}
\end{figure}

Autonomous systems are used in applications such as unmanned aerial and marine vehicles, self-driving cars, smart robots, and many other application areas. 
As shown in Figure \ref{fig:targetdomains}, the space industry dominated by 40\% of the respondents, followed by aviation at 16\%, military at 13\% and automobile at 9\%. Domains included in ``Other'' were financial, government, transportation, Earth Ocean Sciences, and Department of Homeland Security.
Compared to our previous study \cite{goseva2016survey}, which focused only on \MBSE, the number of respondents from the automotive industry in this survey was smaller (i.e., 33\% versus 9\%). 
This might be due to the the proprietary nature of autonomous systems in the automotive industry where respondents are not legally permitted to divulge information.

\subsubsection{RQ1b: What is the level of safety criticality of the applications where autonomous systems and/or MBSwE are being used?}
\begin{figure}[ht]
\begin{center}
\includegraphics[trim=10 200 10 175,clip,width=0.4\textwidth]{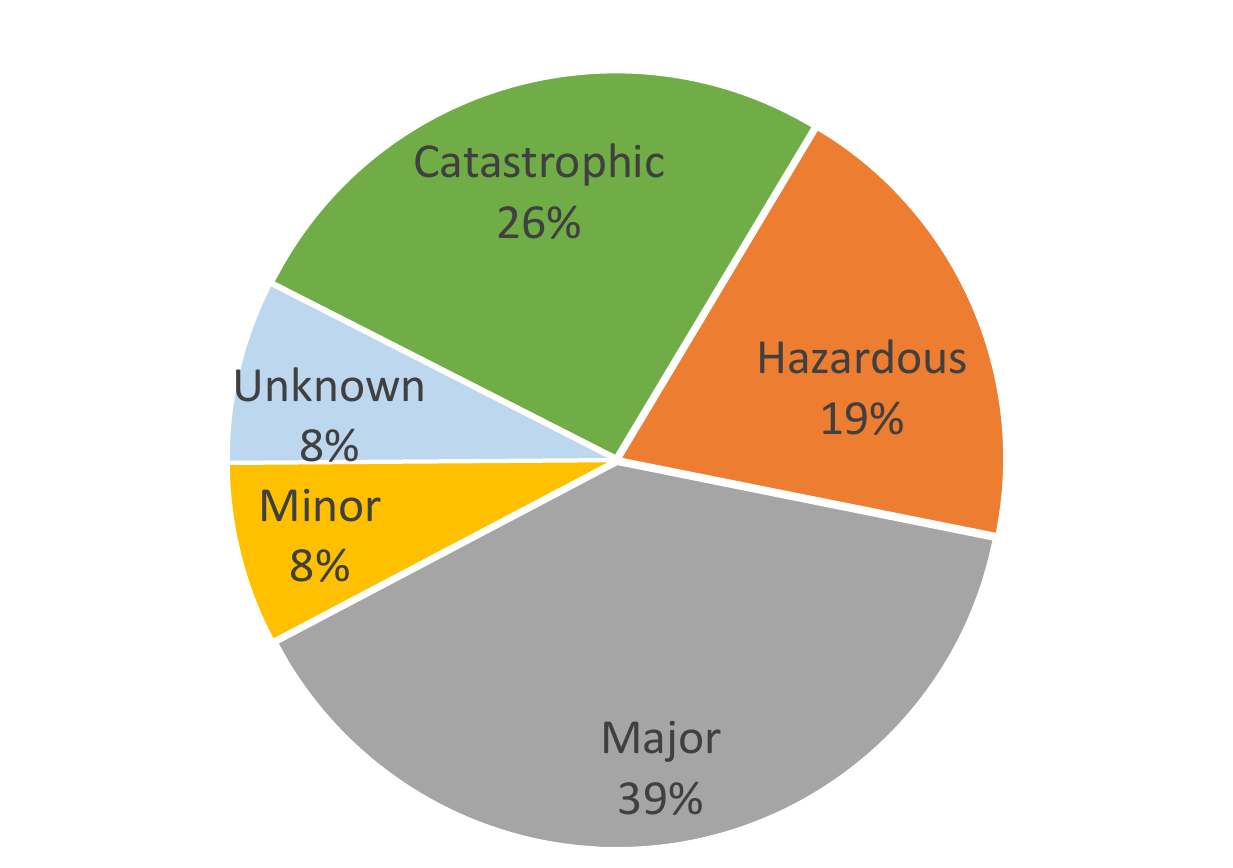}
\end{center}
\caption[Safety criticality of applications]{Safety criticality of applications (97 respondents)}
\label{fig:safetycriticality}
\end{figure}

The level of safety criticality defines the impact of failures if the software or system fails. In the survey, we used the levels of safety criticality as defined in the DO-178C standard \cite{DO-178C}: catastrophic, hazardous, major, minor, and no effect. As shown in Figure \ref{fig:safetycriticality}, 45\% of respondents indicated catastrophic or hazardous level of safety criticality, followed by 39\% with major safety criticality, and 8\% with minor criticality. Only 8\% of respondents indicated an unknown level of safety criticality. 
Since a large portion of the respondents to our survey were from the space and aviation industry and from military, which typically deal with safety-critical applications, it is not surprising that many respondents indicated a high level of safety criticality.
These observations are consistent with the results from our previous survey \cite{goseva2016survey}, which used different values for the levels of criticality but had similar percentage of high criticality of 46\%. 

\subsubsection{RQ1c: What are the programming languages used during the development and deployment of the projects?}
\label{Subsection-RQ1c}
\begin{figure}[ht]
\captionsetup{justification=centering}
\centering
\begin{subfigure}{0.5\textwidth}
    \centering
    \includegraphics[width=0.85\textwidth]{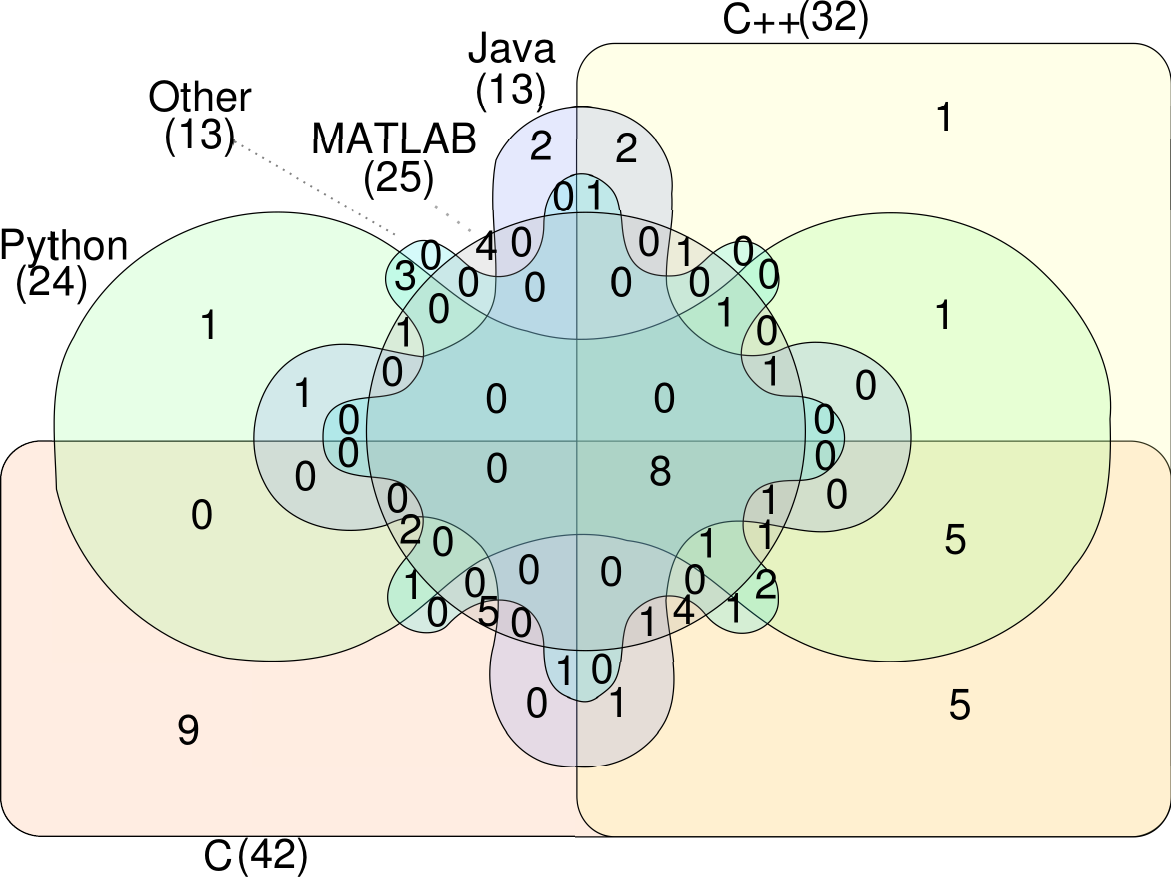}
    \caption[Programming languages used during development]{During development}
    \label{fig:programminglanguagesdevelopment}
\end{subfigure}%
\begin{subfigure}{0.5\textwidth}
    \centering
    \includegraphics[width=0.85\textwidth]{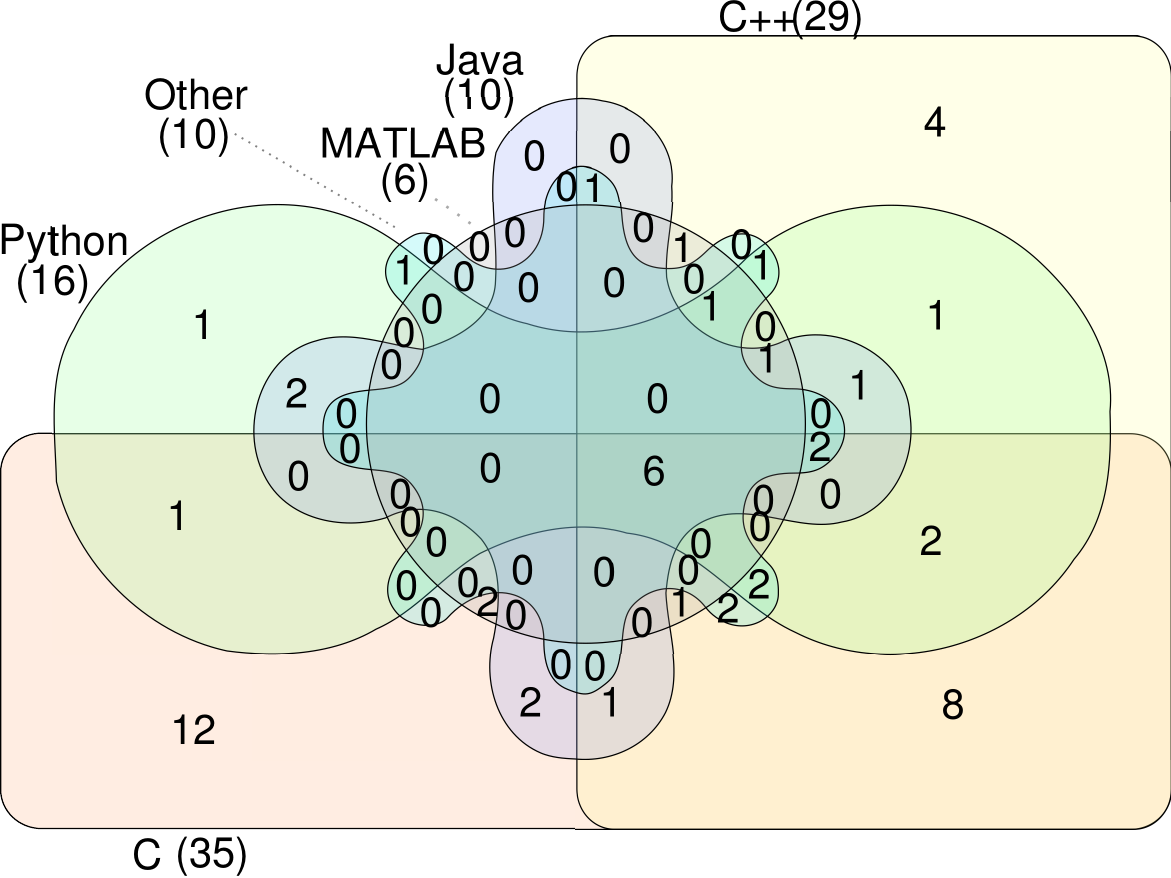}
    \caption[Programming languages used during deployment]{During deployment}
    \label{fig:programminglanguagesdeployment}
\end{subfigure}
\caption{The use of different programming languages (97 respondents, multiple selections possible)}
\end{figure}

In our survey, we distinguished between programming languages and tools that were used during the development and those used in deployed code during operation. Programming languages used during development (e.g., Python or TensorFlow used to train a Deep Neural Network in the cloud) produce intermediate artifacts. Programming languages used in deployed code are executed during system operation. 
Figure \ref{fig:programminglanguagesdevelopment} presents a Venn diagram of the programming languages used during development. Note that the respondents could choose multiple languages. In fact, 8 respondents indicated they used all of the programming languages shown in the figure and most respondents indicated use of multiple languages. C and C++ dominated the responses, with 42 and 32 responses respectively, followed by MATLAB
and Python, 
with 25 and 24 responses, respectively.

Programming languages used during deployment are given in Figure \ref{fig:programminglanguagesdeployment}. Here as well the respondents could choose multiple languages. C and C++ dominated with 35 and 29 responses respectively, followed by Python 
with 16 responses
and MATLAB 
with 6 responses. 
The category of ``Other'' programming languages in Figures \ref{fig:programminglanguagesdevelopment} 
and \ref{fig:programminglanguagesdeployment} included C\#, assembly, Lisp, JavaScript, and G2. 

To compare the use of programming languages during development and deployment, the total number of responses for each language shown in Figures~\ref{fig:programminglanguagesdevelopment} and \ref{fig:programminglanguagesdeployment} are also shown in Table~\ref{table:ProgrammingLanguages}.
These values indicated that the languages used during development and during deployment were very similar. As expected,  embedded AUCs are mainly implemented in C or C++, or a mixture thereof. 
It seems that some languages, which are not very widespread in safety-critical and embedded systems, like Python, are being used as implementation languages for AUCs. Note that there is stronger preference to using Python during development than during deployment. 
The much higher usage of MATLAB during development over deployment 
may be attributed to the use of Mathworks tools for model-based software 
development. Typically, models are developed using Simulink and MATLAB
while the auto-generated code (generated by the Mathworks code generator)
is run during deployment. 
Finally, no arcane or special-purpose languages have been used for development and implementation of AUCs. The most outstanding example here is Lisp.

\begin{table}[th]
\caption[Programming languages used during development and deployment]{Programming languages used during development and deployment (97 respondents, multiple selections possible)}
\label{table:ProgrammingLanguages}
\centering
\begin{tabular}{|l|r|r|r|r|r|r|}
\hline 
& C & C++ & Matlab & Python & Java & Other\\
\hline
Development & 42 & 32 & 25 & 24 & 13 & 13\\
\hline
Deployment & 35 & 29 & 6 & 16 & 10 & 10 \\
\hline
\end{tabular}
\end{table}

\subsubsection{RQ1d: How was the code for autonomous functionality during development and deployment developed?}
\label{subsection:RQ1d}
\begin{figure}[ht]
\captionsetup{justification=centering}
\centering
\begin{subfigure}{0.5\textwidth}
    \centering
    \includegraphics[trim=10 200 10 175,clip,width=0.9\linewidth]{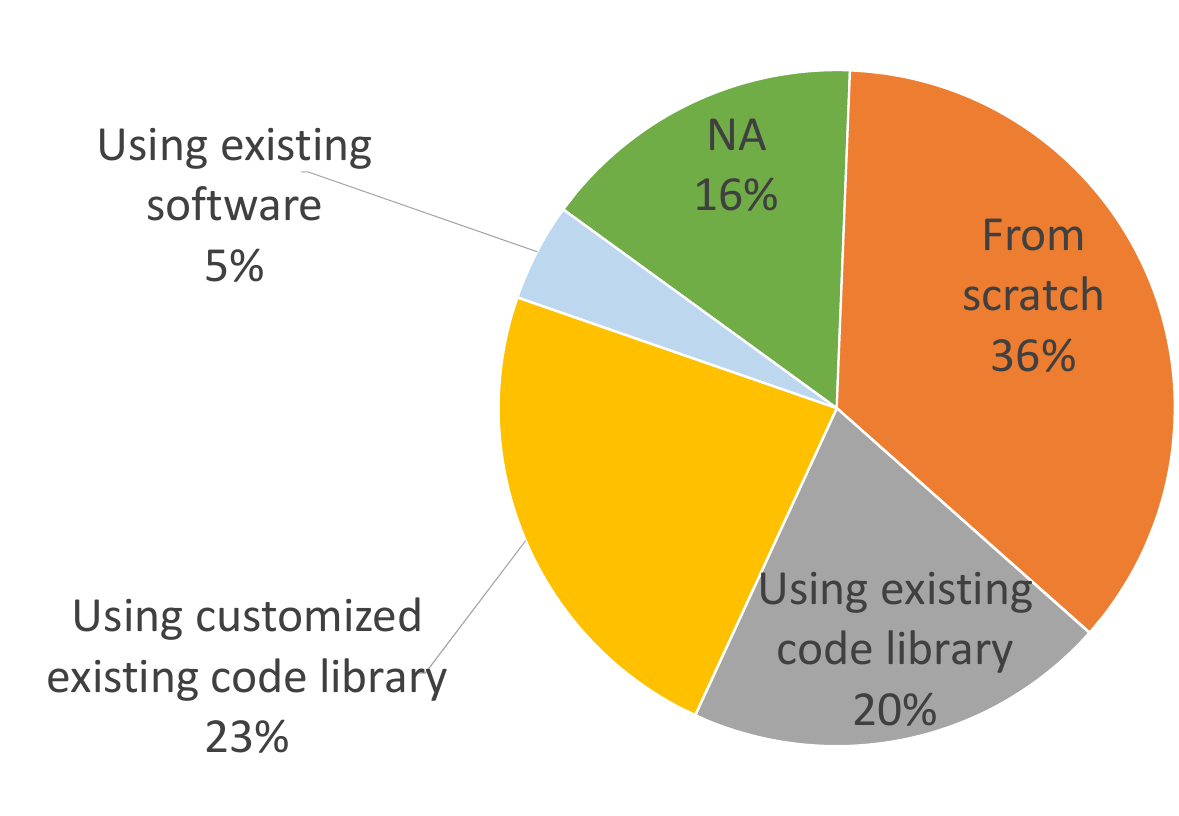}
    \caption[Origins of code during development]{Origins of code during development}
    \label{fig:autonomouscodedevelopment}
\end{subfigure}%
\begin{subfigure}{0.5\textwidth}
    \centering
    \includegraphics[trim=10 200 10 175,clip,width=0.9\linewidth]{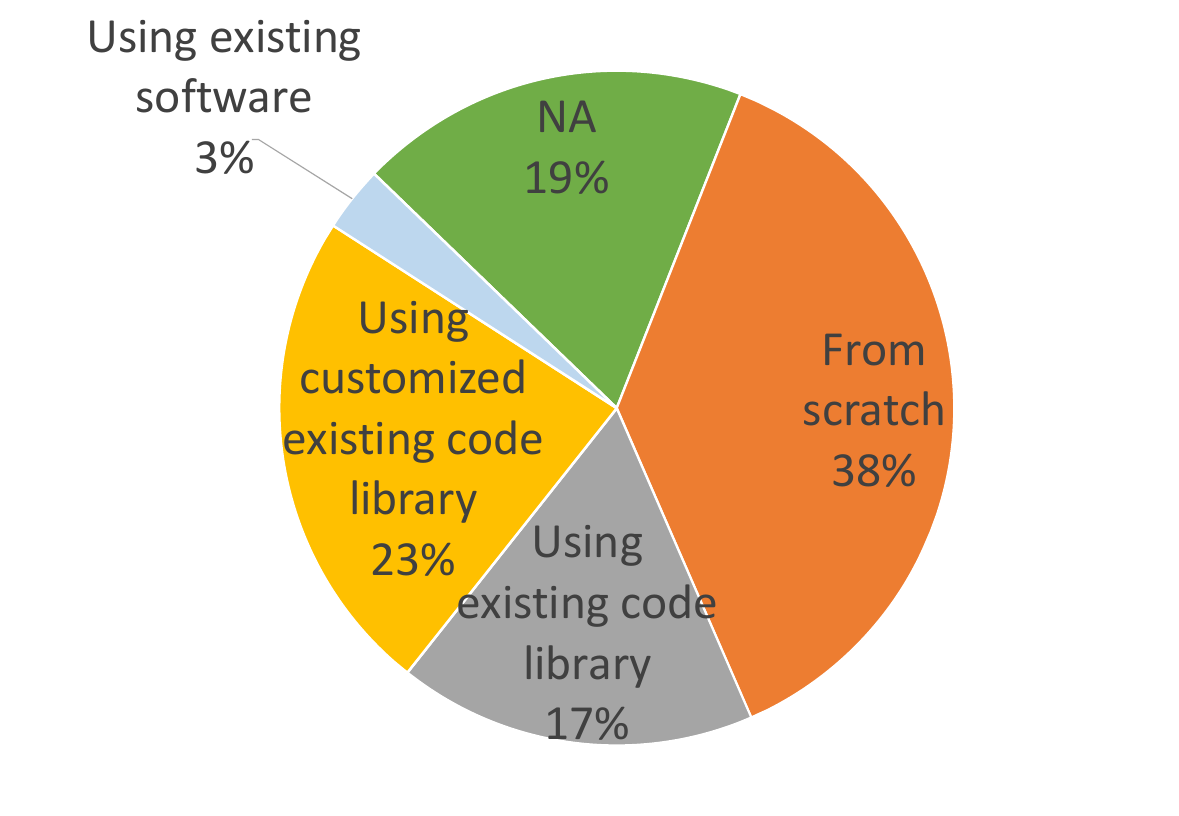}
    \centering
    \caption[Origins of code during deployment]{Origins of code during deployment}
    \label{fig:autonomouscodedeployment}
\end{subfigure}
\label{fig:codeorigins}
\caption{How was the code for autonomous functionality developed (69 respondents)}
\end{figure}

Respondents were asked about how the autonomous code was developed and could only choose one answer from the provided answers in the survey. As can be seen in Figure~\ref{fig:autonomouscodedevelopment}, over one third of the respondents indicated that the code during development was written from scratch, i.e., there was no reuse. 20\% of respondents indicated use of (unmodified) existing code libraries and 23\% of respondents indicated use of customized existing code libraries. Only 5\% of respondents selected reuse of existing software.

Figure \ref{fig:autonomouscodedeployment} depicts the origins of the autonomous code during deployment. (Here also the respondents could choose only one of the options.) The results are similar to those during development, with 38\% of the deployed code developed from scratch, 17\% using unmodified existing code library, 23\% using customized existing code library, and only 3\% reusing existent software. 

\subsubsection{RQ1e: Was special hardware and/or cloud used for developing and running code that has autonomous functionality?}
\label{subsection:RQ1e}

We asked the respondents about the use of special hardware for development and running of the code with autonomous functionality. As shown in Table~\ref{table:hardware}, only 16\% of respondents used special hardware during development and 19\% during operation. The special hardware that was specified by the respondents included touch screens, Nvidia Jetson, and custom designed space robot prototypes.

We also asked if any computations were executed in the cloud or on a remote server. 
Similarly as with the special hardware, the overwhelming majority (i.e., 70\%) of respondents did not execute any software in the cloud or on a remote server. 9\% of respondents used a cloud or remote server only during design, and additional 6\% both during design and operation. (15\% of respondents selected the NA option.) 

The low usage of special hardware and cloud services was somewhat surprising. It may be due to the fact that, as described in RQ2c (Subsubsection~\ref{SubSection-RQ2c}), rule based algorithms still dominated as development approach of autonomy at the time the survey was administered. 

\begin{table}[H]
\begin{center}
\caption[Use of special hardware]{Use of special hardware (69 respondents)}
\label{table:hardware} 
\begin{tabular}{|l|l|l|}
\hline
During:             & development & operation \\ \hline
Special hardware    & 16\%               & 19\%             \\ \hline
No special hardware & 70\%               & 66\%             \\ \hline
NA                  & 14\%               & 15\%            \\ \hline
\end{tabular}
\end{center}
\end{table}

\subsubsection{RQ1f: What were the respondents' roles in the project?}
\begin{figure}[ht]
\begin{center}
\includegraphics[trim=15 260 0 250,clip,width=0.75\textwidth]{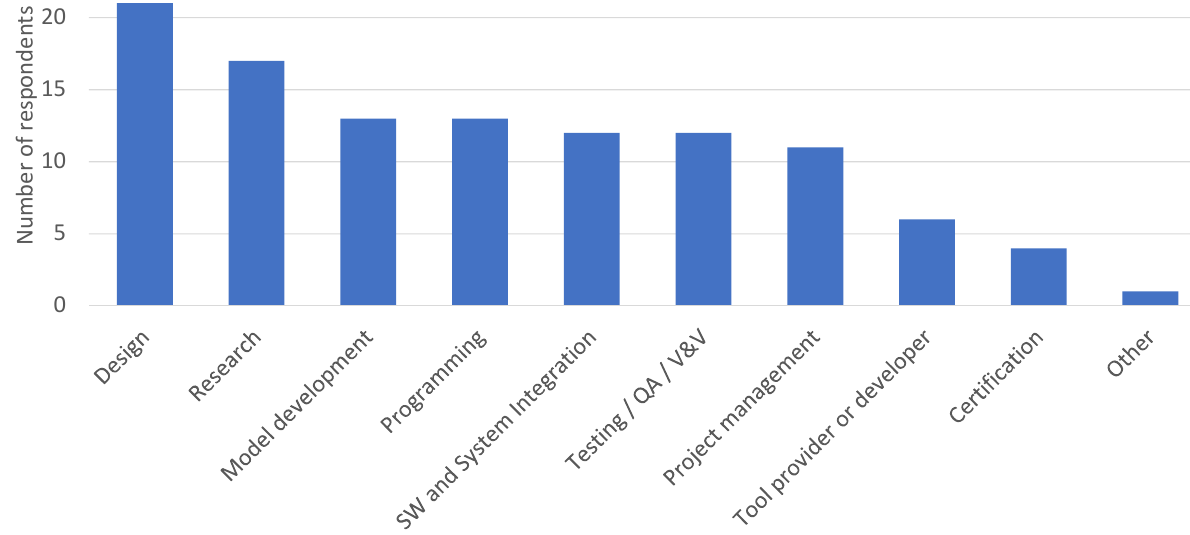}
\end{center}
\caption[Respondents' role in the project]{Respondents' roles in the project (34 respondents, multiple selections possible)}
\label{fig:role}
\end{figure}

Figure \ref{fig:role} presents the respondents' roles in their projects. 
As expected, the two respondents' roles -- Design and Research -- dominated with 21 and 17 respondents, respectively. The roles of Model development, Programming, Software and System integration, Testing/QA/V\&V, and Project Management were roughly evenly distributed with 12 to 13 respondents per category. 
Under the ``Other'' category the respondents mentioned the roles of Certification and Independent Verification and Validation (IV\&V) analyst. Interestingly, no tool developers or vendors were among the respondents, which may be due to the way the survey was distributed. 

\newpage
%%%%%%%%%%%%%%%%%%%%%%%%%%%%%%%%%%%%%%%%%%%%%%%%%%%%%%%%%%%%%%%%%%%%%%%%%%%%%%%%%%%%%%
\subsection{Survey Section 2: Details on Autonomy}
%%%%%%%%%%%%%%%%%%%%%%%%%%%%%%%%%%%%%%%%%%%%%%%%%%%%%%%%%%%%%%%%%%%%%%%%%%%%%%%%%%%%%%

Since an autonomous system can contain more than one \auc, we solicited details about autonomy for each \auc. For this purpose, the respondents were first asked to specify the number of \aucs in the system by choosing a number between 0 and 10. 
The respondents who selected 0 \auc were directed to the reuse section of the survey (see Figure \ref{fig:flowchart}). 
38 respondents who chose one or more \auc were asked to answer a set of questions about autonomy details for each \auc. 
Of these 38 respondents, 28 (i.e., 74\%) provided details for only one \auc. Six respondents provided details about two \aucs each. 
Three sets of two respondents each entered details about three, four and eight \aucs, respectively.  
The analysis presented in this section is based on the data about 58 \aucs entered by 38 respondents. 

\subsubsection{RQ2a: Was \auc developed using \MBSE and which \MBSE tools were used?}

Only 38\% of \aucs were developed using \MBSE. 
Specifically, 17\% of \aucs were developed using MATLAB/Simulink, which explains the relatively high usage of MATLAB during development (see  Figure~\ref{fig:programminglanguagesdevelopment} and Section~\ref{Subsection-RQ1c}). Another 13\% of \aucs were developed using Rational Rhapsody.
Several other less frequently used tools include Rational Rose and Magic Draw. 
The ``Other'' category included the following: Java, Prolog, Papyrus \cite{papyrus}, and Generic Modeling Environment (GME). %need citation

\subsubsection{RQ2b: What was the level of autonomy of \aucs?}

An autonomous component may achieve different levels 
of autonomy. Multiple frameworks that define levels of autonomy have been 
proposed in the past \cite{jahan2019securityLiteratureReview}.    
However, currently no universal framework for defining the levels of autonomy across different domains exists. 
In our survey, for simplicity and to accommodate different application domains, we asked the respondents to select from the following four levels of autonomy: (1) the autonomy is used only as a tool for assistance (i.e., ``the human is primary and the computer is secondary'', (2) ``the computer operates with human interaction'', (3) ``the computer operates independently of the human''
where the human has limited capabilities, and 
(4) ``the computer has full autonomy''.

\begin{figure}[ht]
\begin{center}
\includegraphics[trim=20 260 1 200,clip,width=0.85\textwidth]
{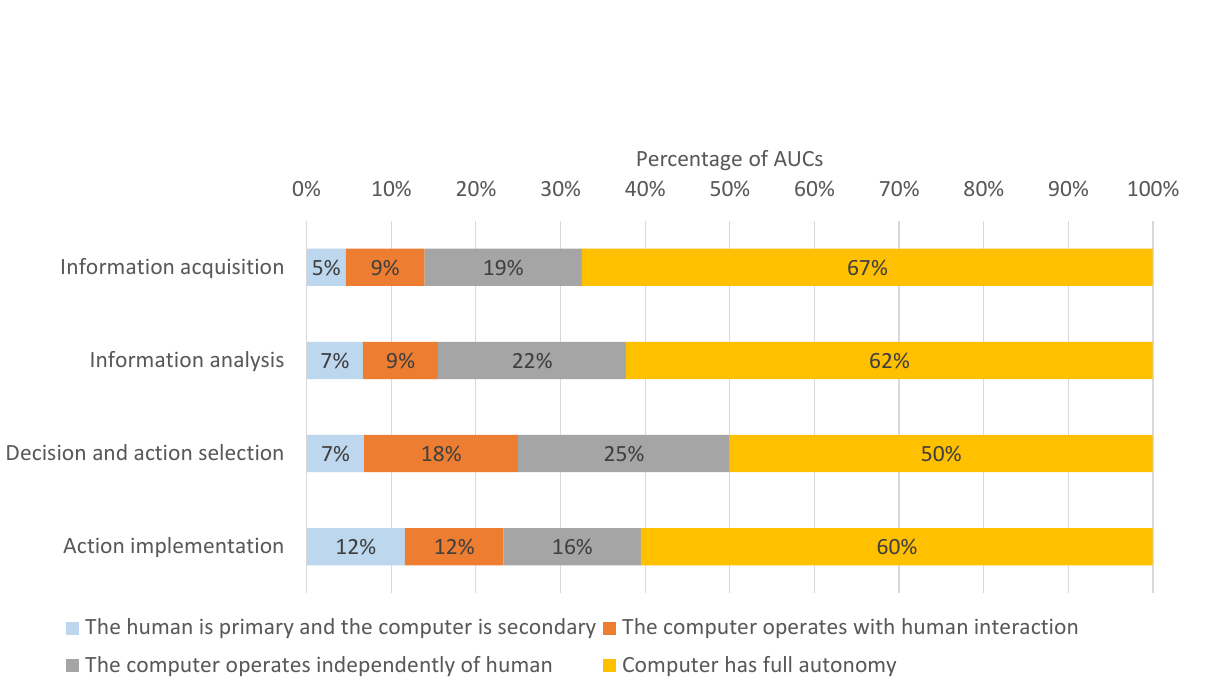}
\end{center}
\caption[Levels of autonomy for specific tasks]{Levels of autonomy for specific tasks (50 \aucs)}
\label{fig:auctaskautonomy}
\end{figure}

Following \cite{parasuraman2000model}, we included the following 
autonomy tasks in the survey:%
~Information acquisition (monitor),
Information analysis (analyze),
Decision and action selection (decide), and
Action implementation (act). 
Figure \ref{fig:auctaskautonomy} presents the results for the level of autonomy for these four tasks, for 50 \aucs. Interestingly, for each task, a significant percentage of \aucs (i.e., between 50\% and 67\%) operated with the computer having a full autonomy. 
Of the four tasks, ``Decision and action selection'' had the highest dependence on human interaction, followed by ``Action implementation''.
Only 5\% to 12\% of \aucs had the lowest level of autonomy (i.e., ``the human primary and the computer secondary'') for different tasks. 

It appears that the level of autonomy has increased significantly when compared to the results reported in the related survey \cite{lavallee2006intelligent}, which focused on autonomous systems from the space domain and concluded that the lower levels of autonomy dominated (with 45\% of implementations having only automatic notification). 

\subsubsection{RQ2c: What algorithms and modeling paradigms were used to develop \aucs?}
\label{SubSection-RQ2c}
\begin{figure}[ht]
\begin{center}
\includegraphics[trim=20 225 0 200,clip,width=0.7\textwidth]{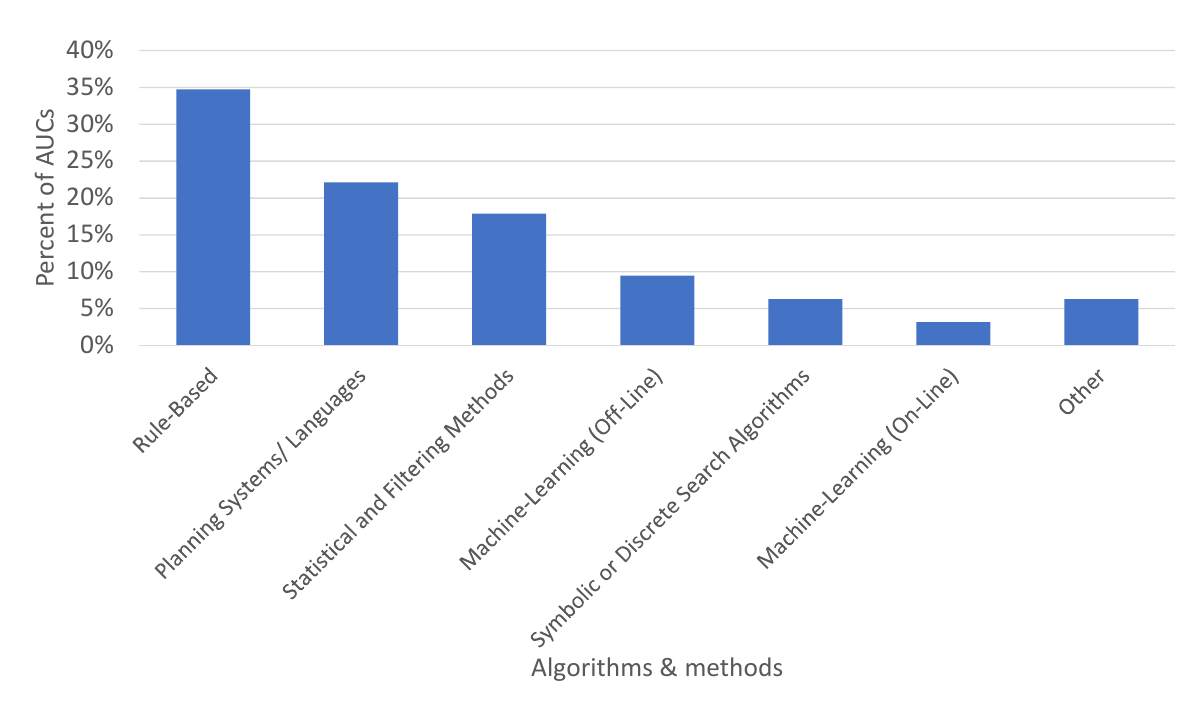}
\end{center}
\caption[Algorithms and modeling paradigms for autonomous components]{Algorithms and modeling paradigms used for development of \aucs (58 \aucs, multiple selections possible)}
\label{fig:aucmodelingparadigms}
\end{figure}

As can be seen in Figure \ref{fig:aucmodelingparadigms}, which shows the percentage of \aucs developed using different algorithms and modeling paradigms;
``Rule-Based'' algorithms and methods, which are based on program logic or using explicit rules, were used for development of 35\% of \aucs.  
``Planning Systems/Languages'' and ``Statistical and filtering methods'' were used for development of 22\% and 18\% of \aucs, respectively. 
Interestingly, at the time our survey was conducted (i.e., mid 2019) these traditional algorithm and methods dominated the development of \aucs.  
We were surprised that machine learning approaches, which at that time were much hyped by the media as the core for autonomy, were used much less frequently (i.e., only 12\%). Specifically, only 9\% of \aucs used offline machine learning and 3\% used online machine learning.
(The ``Other'' category included nonlinear programming techniques, model checking, and nonlinear optimization.)

The low usage of machine learning approaches explains the finding of RQ1d (Section~\ref{subsection:RQ1d}) related to the high percentage of code developed from scratch. 
Namely, higher usage of machine learning may have led to use of 
off-the-shelf algorithms for data preparation and training 
(e.g., TensorFlow, Keras, etc.) and thus would have increased the percentage of reuse. 
Furthermore, the low usage of machine learning approaches may be the reason 
behind the low usage of special hardware and cloud services as 
found in RQ1e (Section~\ref{subsection:RQ1e}).  

\subsubsection{RQ2d: How were the requirements for the \aucs specified?}

\begin{figure}[th]
\centering
\begin{minipage}[b]{0.45\textwidth}
    \centering
    \includegraphics[scale=0.3]{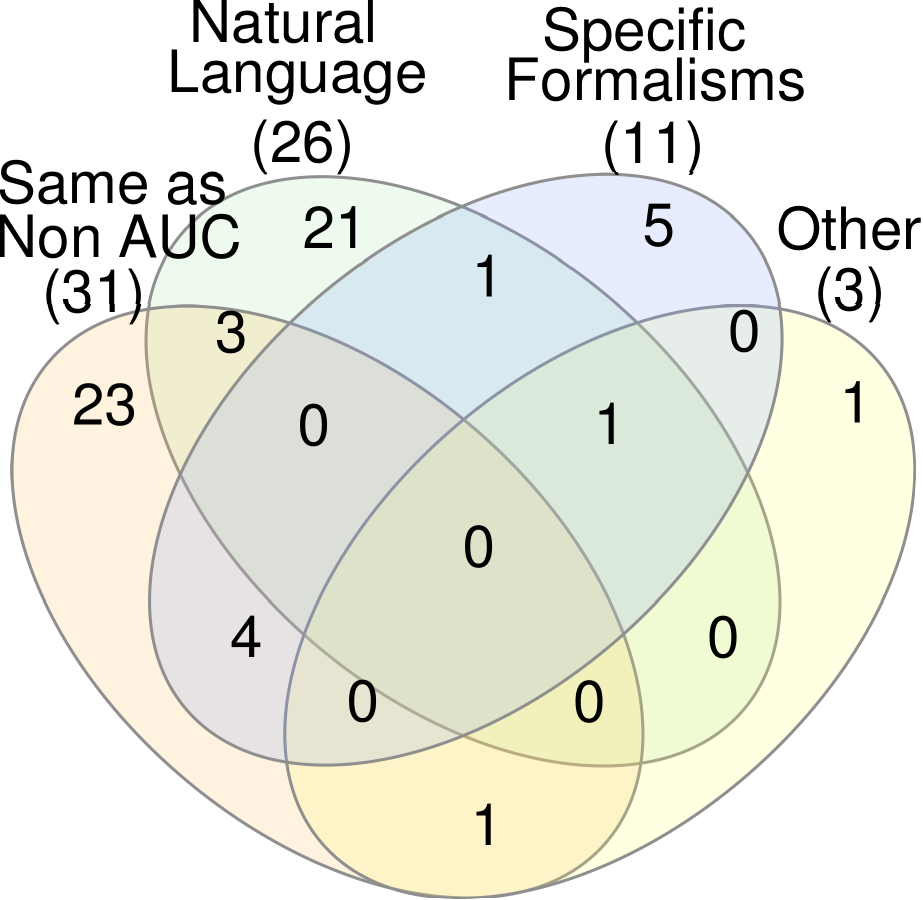}
    \caption[Specification of requirements for \aucs]{Specification of requirements for \auc{s} (51 \auc, multiple selections possible)}
    \label{fig:aucrequirements}
\end{minipage}%
\hfill
\begin{minipage}[b]{0.45\textwidth}
    \centering
    \includegraphics[trim=10 200 10 200,clip,width=0.99\textwidth]{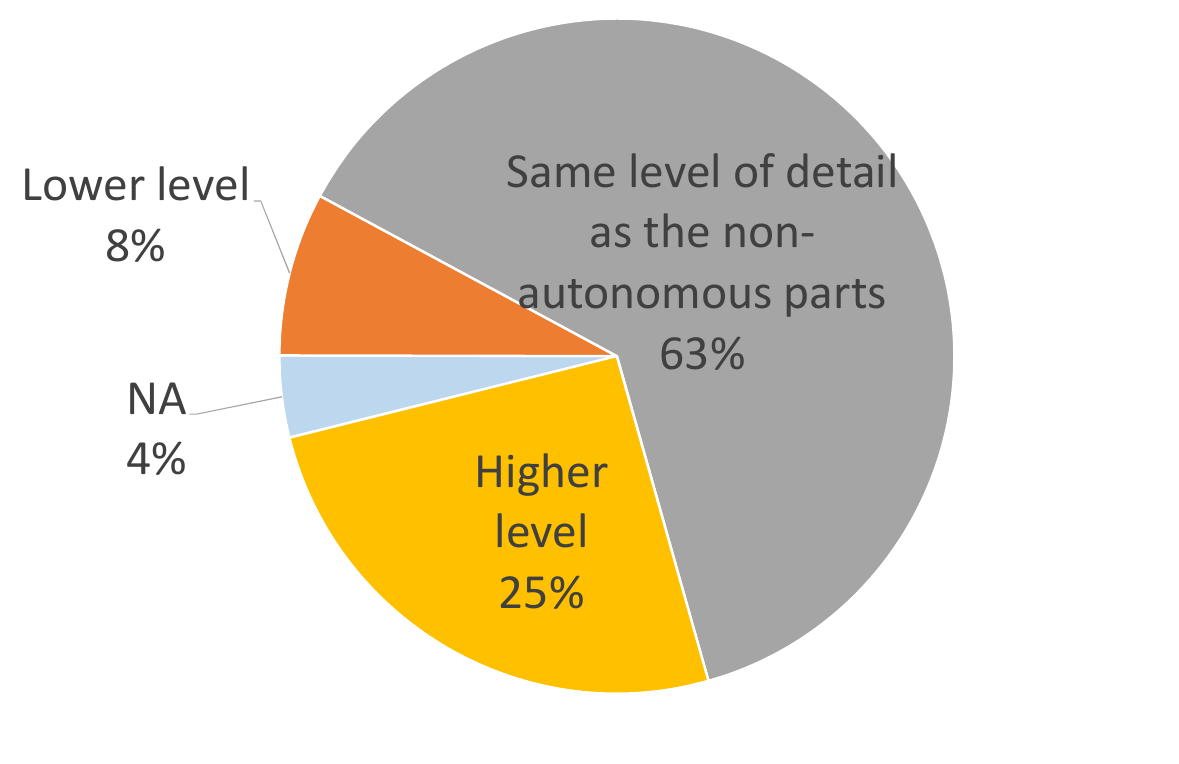}
    \caption[Level of details of requirements specification for \aucs compared to non-\aucs]{Level of details of requirements specification for \aucs compared to non-\aucs (51 \auc)}
    \label{fig:aucrequirementscomparedtononauc}
\end{minipage}
\end{figure}

We also explored the ways requirements for \aucs were specified. The respondents could select more than one option for requirements specification of each \auc.
As can be seen in Figure~\ref{fig:aucrequirements}, only 10 \aucs were developed using more than one approach for requirement specification. 
For 31 \aucs, the requirements were specified same way as for non-\aucs. Natural Language was used more than twice as often than some form of formal specification.     

The survey also included a question on the level of details of requirement specification for \aucs compared to non-\aucs. (For this question, the answer choices were mutually exclusive.) 
The majority of \aucs (i.e., 63\%) were developed using requirements with 
the same level of details as the non-\aucs 
(Figure \ref{fig:aucrequirementscomparedtononauc}).
However, for a quarter of \aucs the requirements specification was at higher level of details than for non-\aucs. 

\subsubsection{RQ2e: What were the challenges associated with the \aucs?}
\begin{figure}[t]
\begin{center}
\includegraphics[trim=20 255 0 225,clip,width=0.8\textwidth]{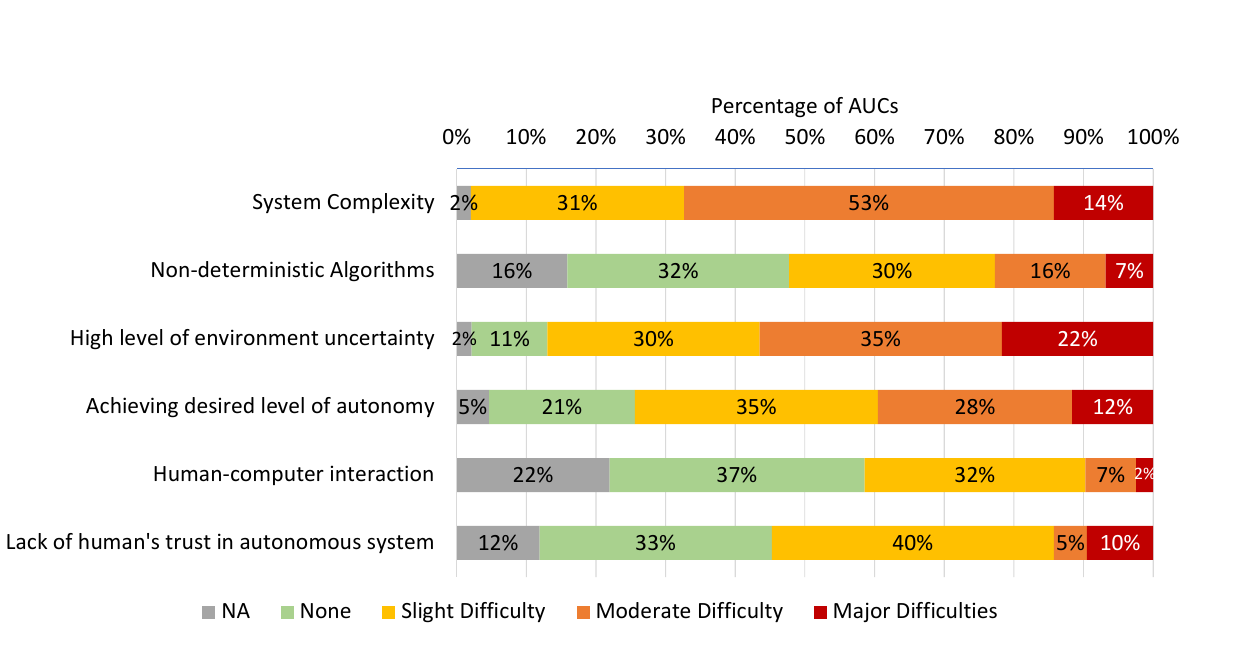}
\end{center}
\caption{Degree of difficulty for challenges encountered during development, deployment and use of \aucs (50 \aucs)}
\label{fig:aucdifficulties}
\end{figure}

We also explored the challenges associated with \aucs, asking the respondents to use an ordinal scale for six different challenges given in Figure~\ref{fig:aucdifficulties}. (Since the numbers of responses were not the same for all six challenges, to allow for comparison, Figure \ref{fig:aucdifficulties} shows the percentages of the degrees of difficulty for each challenge.)
If we consider the moderate and major difficulties together, ``System complexity'' was the most challenging (in 67\% of the \aucs), followed by ``High level of environment uncertainty'' and ``Achieving the desired level of autonomy'', in 57\% and 40\% of the \aucs. respectively.
While ``Non-deterministic algorithms'' led to some difficulties (in 23\% of \aucs), ``Lack of human's trust in autonomous systems'' and ``Human-computer interaction'' did not seem to be big issues (affecting only 15\% and 9\% of \aucs, respectively).

%\FloatBarrier
%%%%%%%%%%%%%%%%%%%%%%%%%%%%%%%%%%%%%%%%%%%%%%%%%%%%%%%%%%%%%%%%%%%%%%%%%%%%%%%%%%%%%%
\subsection{Survey Section 3: Details on Reuse of Software Artifacts}
%%%%%%%%%%%%%%%%%%%%%%%%%%%%%%%%%%%%%%%%%%%%%%%%%%%%%%%%%%%%%%%%%%%%%%%%%%%%%%%%%%%%%%

For the reuse of software artifacts, we explored reuse for both \aucs and non-\aucs. 
As can be seen from the survey flowchart in Figure \ref{fig:flowchart}, only respondents who answered ``yes'' to the question ``Were software, data, or other artifacts reused in your project(s)?'' were  presented with the survey questions related to reuse (i.e., Survey Section 3). 

\subsubsection{RQ3a: What artifacts were reused and to what extent?}
\begin{figure}[th]
\begin{center}
\includegraphics[trim= 60 255 55 225,clip,width=0.8\textwidth]{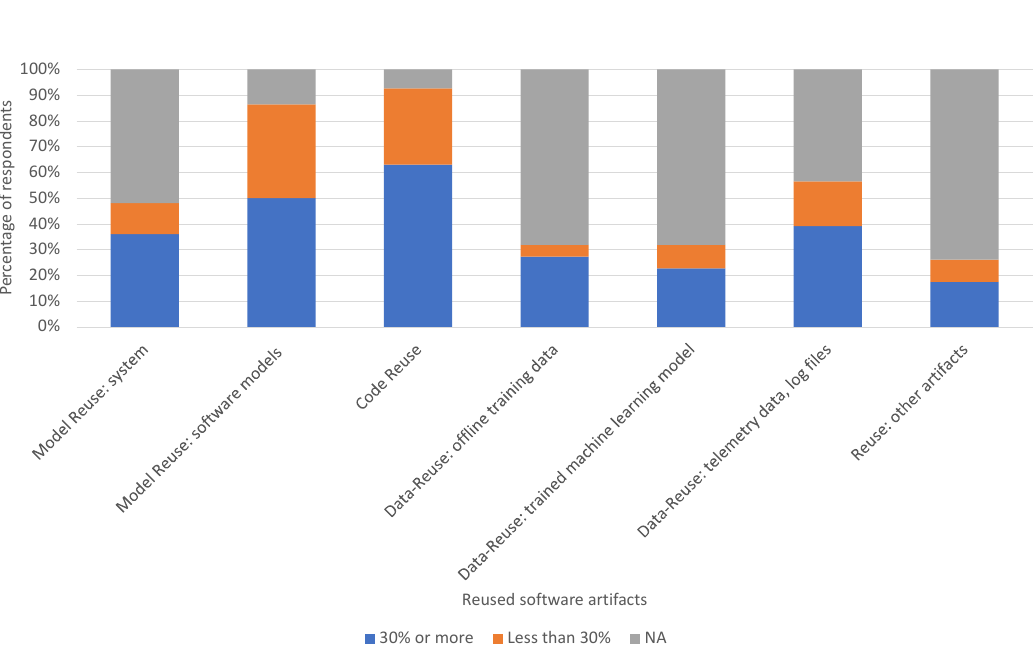}
\end{center}
\caption[The extent of reuse of different software artifacts for \aucs ]{The extent of reuse of different software artifacts for \aucs (30 respondents)}
\label{fig:reuseartifactsaucstackedbar}
\begin{center}
\includegraphics[trim= 60 255 55 225,clip,width=0.8\textwidth]{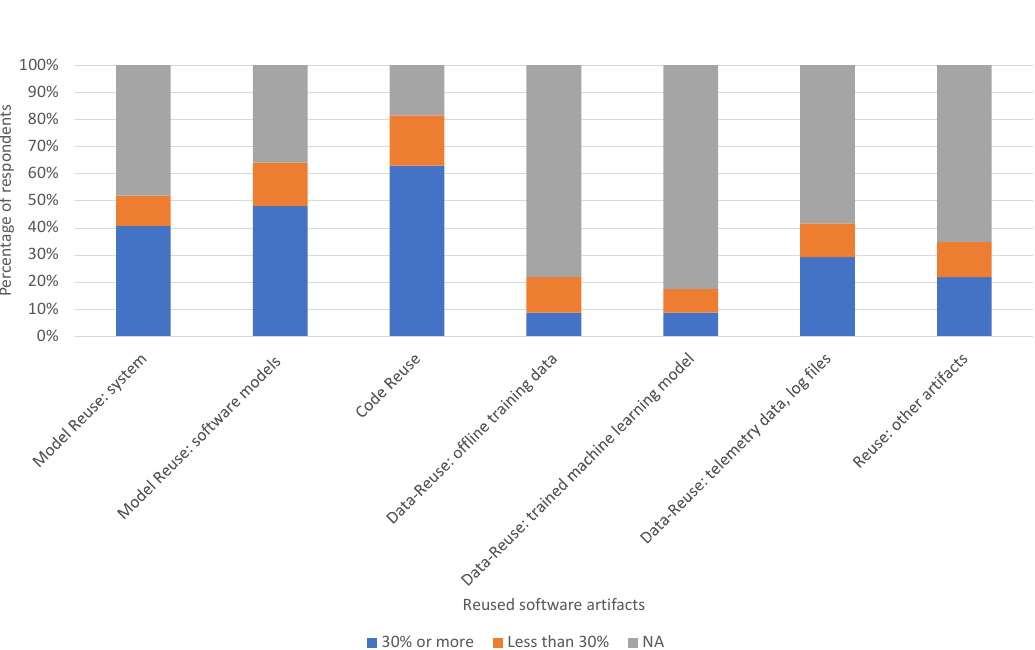}
\end{center}
\caption[The extent of reuse of different software artifacts for non-\auc ]{The extent of reuse of different software artifacts for non-\auc (30 respondents)}
\label{fig:reuseartifactsnonaucstackedbar}
\end{figure}

Figures  \ref{fig:reuseartifactsaucstackedbar} and \ref{fig:reuseartifactsnonaucstackedbar}
show the amount of reuse for different types of artifacts for \aucs and non-\aucs, respectively. 
For each type of software artifacts, respondents could select 30\% or more reuse, less than 30\% reuse, and Not Applicable (NA). 

As can be seen in Figure~\ref{fig:reuseartifactsaucstackedbar},
in case of \aucs, the code had the highest amount of reuse with 63\% of the respondents' projects reusing 30\% or more of the code, followed by 50\% of respondents reusing more than 30\% of ``Model reuse: software models'', and 39\% of respondents having 30\% or more ``Data reuse: telemetry data and log files''. 
When comparing the reuse of software artifacts in \aucs and non-\aucs (see Figures \ref{fig:reuseartifactsaucstackedbar} and \ref{fig:reuseartifactsnonaucstackedbar}) similar trends were observed for all type of software artifacts except ``Data reuse: offline training data'' and ``Data reuse: training machine learning model''. As expected, these two artifacts have much less reuse in case of non-\aucs than in \aucs.  

\subsubsection{RQ3b: Are there any negative aspects of reuse?}
\begin{figure}[t]
\begin{center}
\includegraphics[trim=10 300 0 250,clip,width=0.8\textwidth]{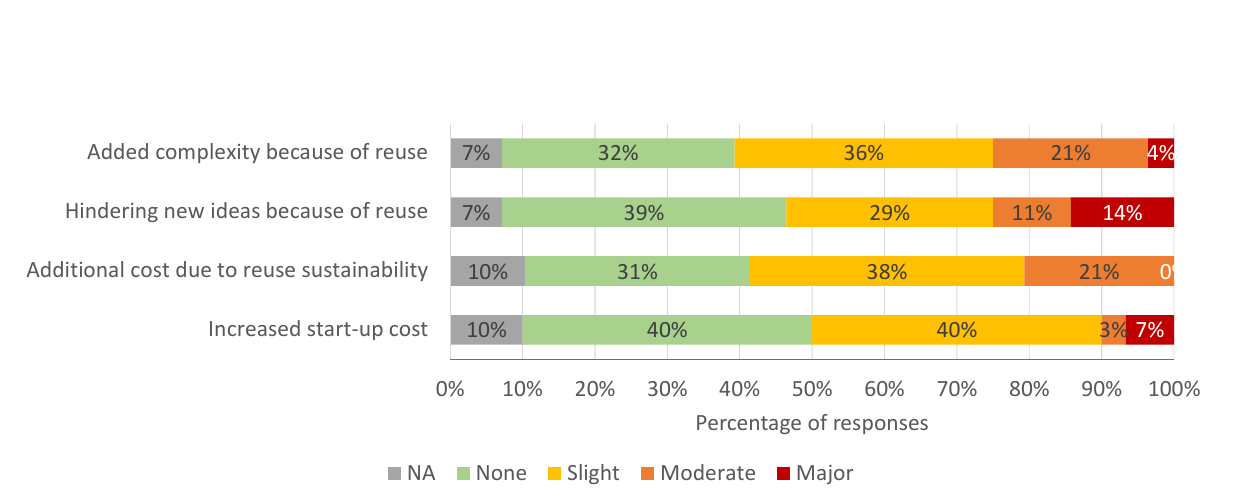}
\end{center}
\caption[Negative aspects of reuse]{Negative aspects of reuse (30 respondents)}
\label{fig:negativeaspectsreuse}
\end{figure}

For this research question we asked the respondents to rate (using an ordinal scale) four different negative aspects of reuse. Because it was not required to respond to each aspect, the numbers of responses for each aspect were slightly different.  
Therefore, Figure \ref{fig:negativeaspectsreuse} shows the percentages of respondents. 
As seen in Figure \ref{fig:negativeaspectsreuse}, from 31\% to 40\% of respondents did not observe any negative aspects due to reuse. The largest negative aspect of reuse was due to ``Hindering new ideas'' with 14\% of respondents selecting major and 11\% selecting moderate negative aspect. This was followed by ``Added complexity because of reuse'' (with 4\% major and 21\% moderate) and ``Additional cost due to reuse sustainability'' (with 21\% moderate negative aspects). 
Under ``Other'' negative aspects, respondents listed lack of documentation and difficulty adding patches into a formal design process.

\subsubsection{RQ3c: What were the difficulties due to reuse?}
\begin{figure}[th]
\begin{center}
\includegraphics[trim=10 270 0 240,clip,width=0.8\textwidth]{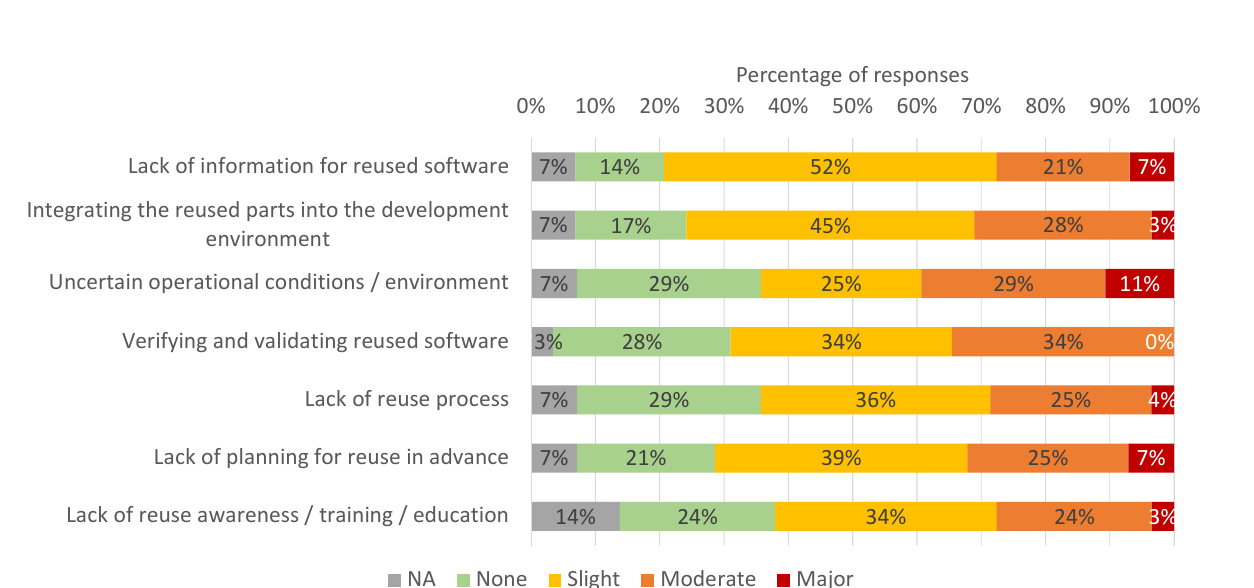}
\end{center}
\caption[Difficulties due to reuse]{Difficulties due to reuse (29 respondents)}
\label{fig:reusedifficulties}
\end{figure}

The amount of difficulty in different aspects of reuse are shown in Figure~ \ref{fig:reusedifficulties}, using an ordinal scale. 
As seen in Figure~\ref{fig:reusedifficulties}, ``Uncertain operational conditions / environment'' led to most difficulties, with 11\% of respondents indicating major and 29\% indicating moderate difficulties. ``Lack of planning for reuse in advance'' and ``Lack of information for reused software'' each had 7\% of respondents selecting major difficulties and 25\% and 21\% moderate difficulties, respectively. 

\subsubsection{RQ3d: What were the benefits of reuse?}
\begin{figure}[ht]
\centering
\begin{subfigure}[b]{0.50\textwidth}
    \centering
    \includegraphics[trim=25 300 5 300,clip,width=1\linewidth]{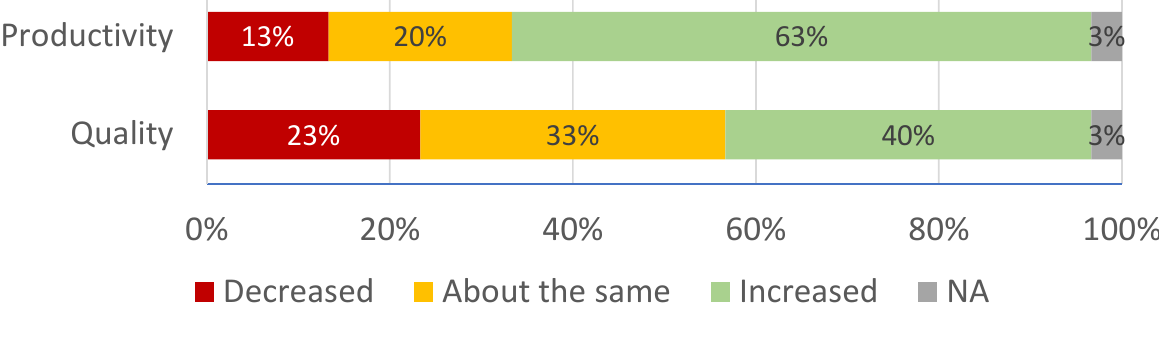}
\end{subfigure}%

\begin{subfigure}[b]{0.50\textwidth}
    \centering
    \includegraphics[trim=30 300 5 300,clip,width=1\linewidth]{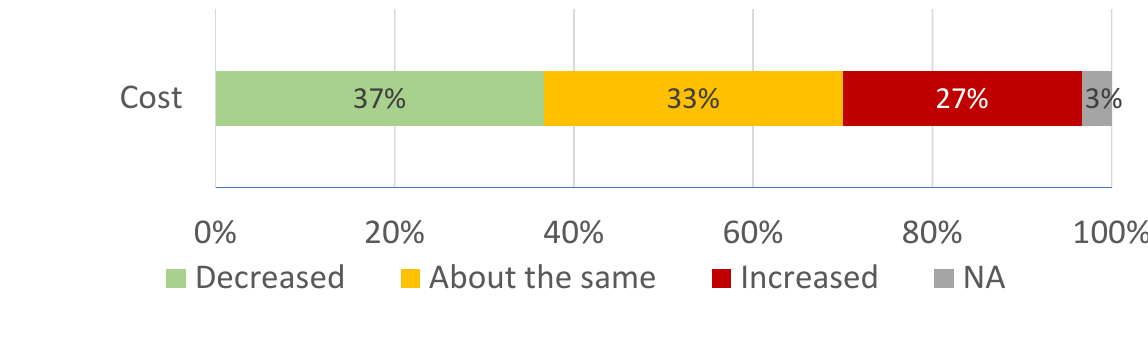}
\end{subfigure}
\caption{Benefits of reuse with respect to productivity, quality, and cost (30 respondents)}
\label{fig:reusebenefits}
\end{figure}

Figure \ref{fig:reusebenefits} shows the benefits of reuse, in terms of productivity, quality, and cost. As can be seen, 63\% of respondents reported increased productivity and 40\% reported increased quality due to reuse. Interestingly, 23\% of respondents reported decreased quality due to reuse. 
Surprisingly, only 37\% of respondents reported decreased cost due to reuse. Note that the distribution was fairly uniform among ``decreased'', ``about the same'', and `'increased'' cost due to reuse. 

%\FloatBarrier
%%%%%%%%%%%%%%%%%%%%%%%%%%%%%%%%%%%%%%%%%%%%%%%%%%%%%%%%%%%%%%%%%%%%%%%%%%%%%%%%%%%%%%
\subsection{Survey Section 4: Processes and Standards}
%%%%%%%%%%%%%%%%%%%%%%%%%%%%%%%%%%%%%%%%%%%%%%%%%%%%%%%%%%%%%%%%%%%%%%%%%%%%%%%%%%%%%%
In this section we examine the processes and standards used when developing systems that have autonomous functionality and/or utilized model-based software development. 
As can be seen in the survey flowchart shown in Figure~\ref{fig:flowchart}, all respondents of the survey were presented with the questions from Section~4 of the survey. However, since this section was towards the end of the survey, some respondents did not complete this section. 

\subsubsection{RQ4a: Which life-cycle model was used?}
\begin{figure}[ht]
\begin{center}
\includegraphics[trim=10 200 0 170,clip,width=0.45\textwidth]{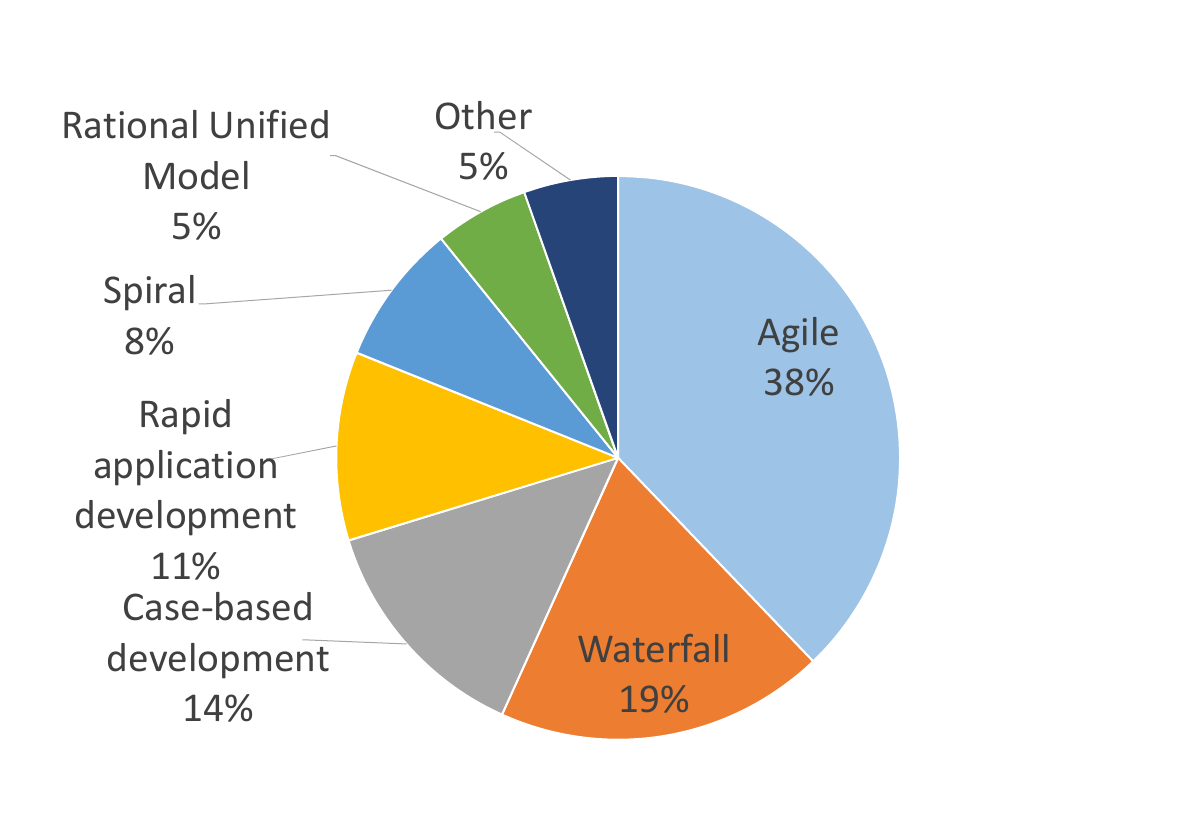}
\end{center}
\caption[Life-cycle model used by projects]{Life-cycle models used by projects (39 respondents)}
\label{fig:lifecycle}
\end{figure}

Figure \ref{fig:lifecycle} shows different types of life-cycle models used on projects. Interestingly, the Agile life-cycle model was used twice as often than the Waterfall model (i.e., 38\% vs.\ 19\%). Case-based development was used by 14\% of respondents, followed by Rapid application development used by 11\%, Spiral  used by 8\%, and Rational Unified Model used by 5\% of respondents. 

\subsubsection{RQ4b: Which modeling standards and coding standards were used by the projects?}
\begin{figure}[ht]
\centering
\begin{minipage}[b]{0.4\textwidth}
    \centering
    \includegraphics[trim=10 200 10 150,clip,width=\textwidth]{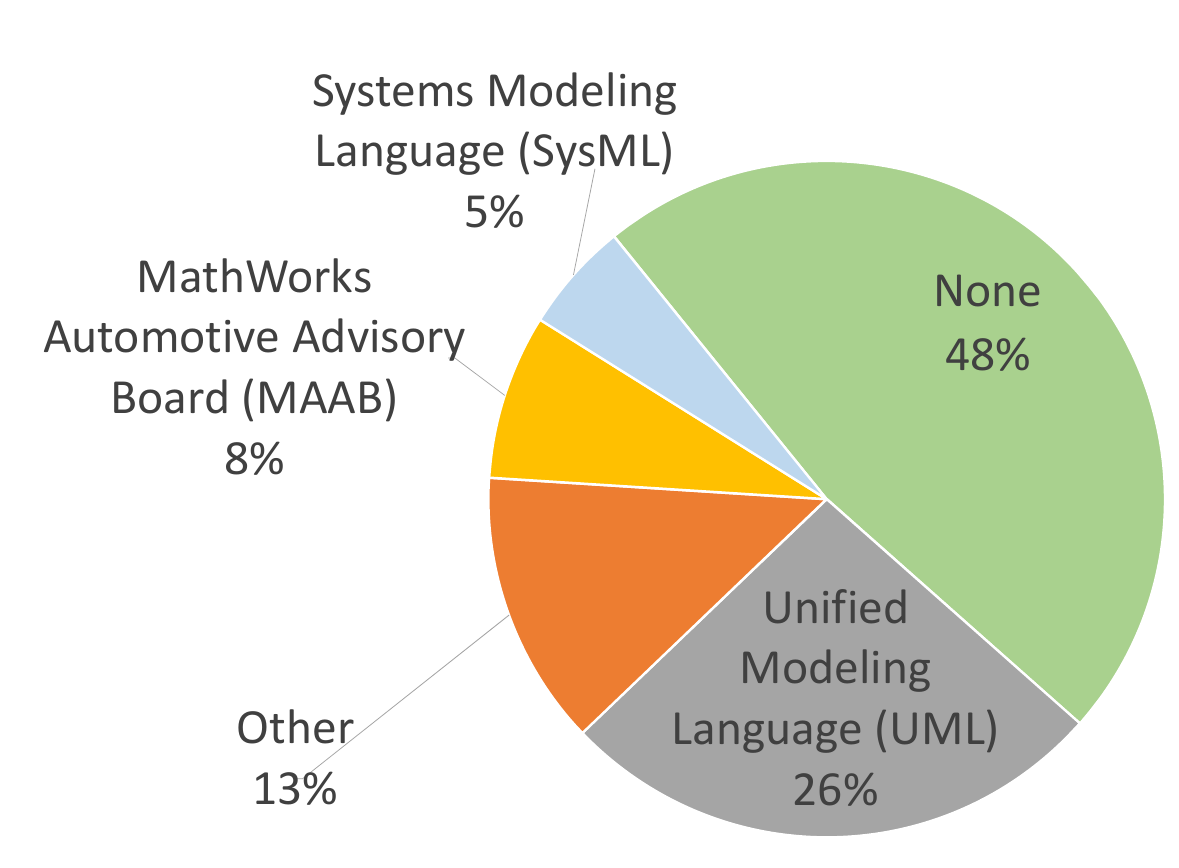}
    \caption[Modeling Standard of Project]{Modeling standards used by projects\\ (40 respondents)}
    \label{fig:modelingstandards}
\end{minipage}%
\hfill
\begin{minipage}[b]{0.4\textwidth}
    \centering
    \includegraphics[trim=10 200 10 150,clip,width=0.9\textwidth]{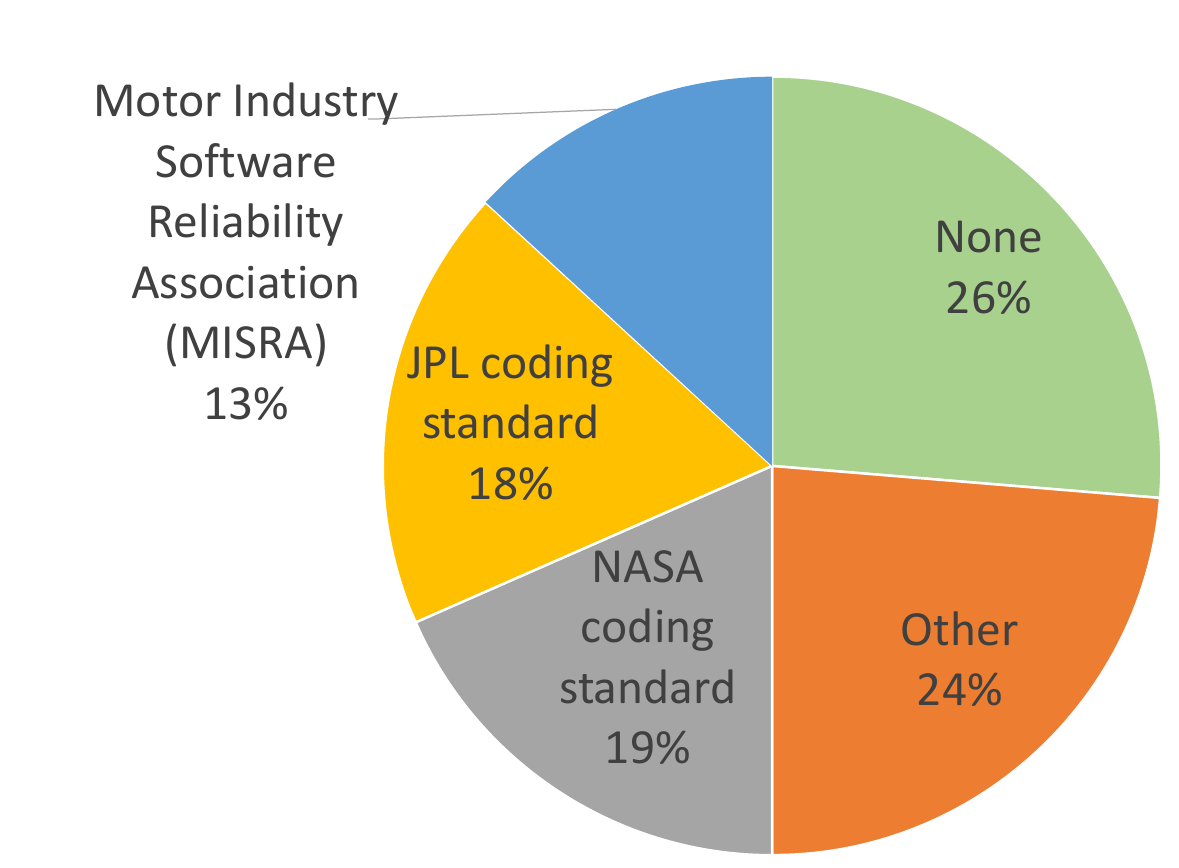}
    \caption[Coding Standards of Project]{Coding standards used by projects\\ (40 respondents)}
    \label{fig:codingstandards}
\end{minipage}
\end{figure}

As shown in Figure \ref{fig:modelingstandards}, almost half of the respondents (i.e., 48\%) did not use any modeling standard. 
The most widely used standard was UML
by 26\% of respondents, followed by %\mathworks 
MathWorks MAAB \cite{MAAB} by 8\% and SysML \cite{SysML} by 5\% of the respondents. 

As can be seen in Figure \ref{fig:codingstandards}, which depicts the coding standards used by the projects, about a quarter of the respondents did not follow any coding standard. Another quarter used standards not explicitly listed in the survey (i.e., ``Other''), which included responses such as Orion, Thales, and internal coding standard. Large percentage of respondents used the NASA (19\%) and JPL (18\%) coding standards, followed by 13\% who used MISRA coding standards.

\subsubsection{RQ4c: Did the system go through a certification process and was the \auc part of the certified system?}

As can be seen in Table \ref{table:additionalaucinfo}, only 24\% of systems went through certification and only 26\% of \aucs were part of a certified system. The respondents who indicated their projects went through a certification process listed the following standards: NASA, Security Technical Implementation Guide (STIG), %need STIG citation?
or an internal standard. 
\begin{table}[ht]
\begin{center}
\caption[Information on certification process]{Information on certification process (41 respondents)}
\label{table:additionalaucinfo}
\begin{tabular}{|l|l|l|}
\hline
                                                    & Yes  & No   \\ \hline
System went through certification process           & 24\% & 76\% \\ %\hline
Autonomous components were part of a certified system & 26\% & 74\% \\ \hline
\end{tabular}

\end{center}
\end{table}

%\FloatBarrier
%%%%%%%%%%%%%%%%%%%%%%%%%%%%%%%%%%%%%%%%%%%%%%%%%%%%%%%%%%%%%%%%%%%%%%%%%%%%%%%%%%%%%%
\subsection{Survey Section 5: Verification and Validation}
%%%%%%%%%%%%%%%%%%%%%%%%%%%%%%%%%%%%%%%%%%%%%%%%%%%%%%%%%%%%%%%%%%%%%%%%%%%%%%%%%%%%%%

All respondents of the survey were presented with the questions from Section 5 of the survey (see the survey flowchart in Figure \ref{fig:flowchart}). However, since this section was towards the end of the survey, some respondents did not complete it. 

\subsubsection{RQ5a: Which quality attributes were verified and validated?}
The survey had a question on verification and validation of each of the following quality attributes: correctness, performance, robustness, safety, and security. Respondents were allowed to select all that applied.
The performance and correctness of the software were verified and validated most frequently, by 83\% and 78\% of the 36 respondents who answered this question, respectively. 
Smaller percentages of respondents verified and validated the robustness and safety (i.e., 61\% and 56\%, respectively). 
Surprisingly, only 11\% of respondents verified and validated to security; these respondents verified and validated all the other quality attributes as well. 

\subsubsection{RQ5b: How were the models verified and validated?}
\begin{figure}[!ht]
\begin{center}
\includegraphics[trim=20 235 0 210,clip,width=0.68\textwidth]{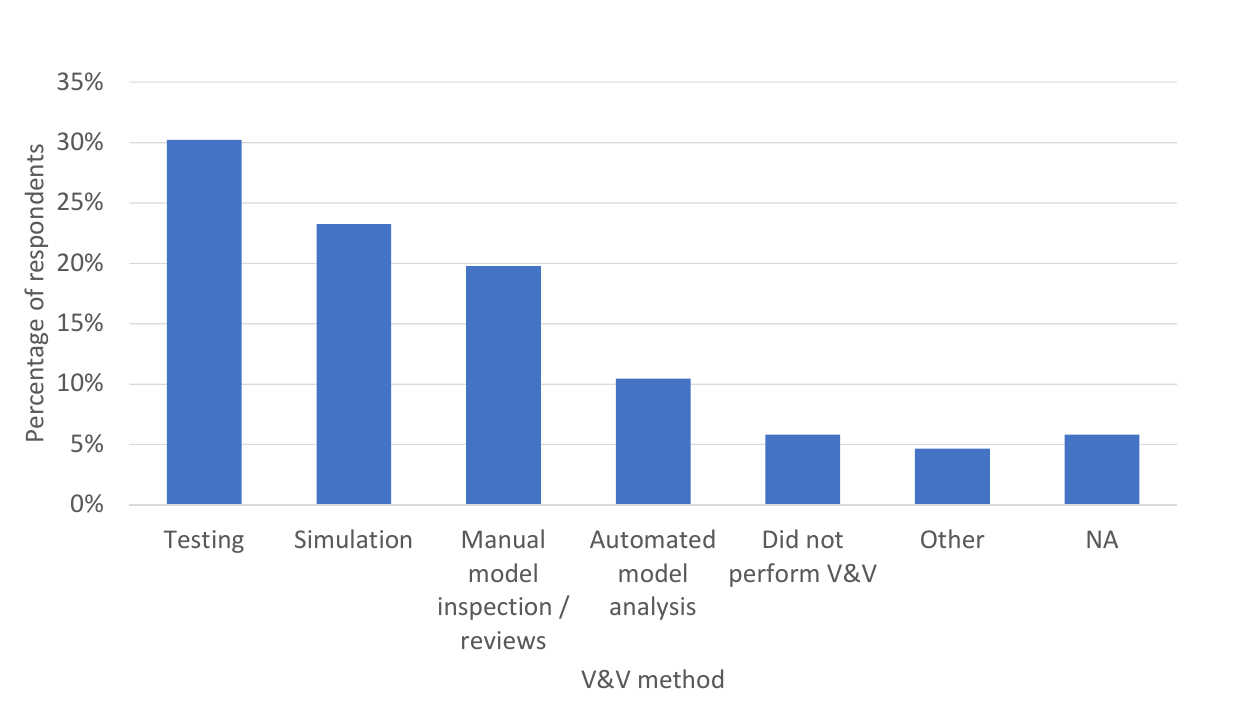}
\end{center}
\caption{Methods used for verification and validation of the models (37 respondents, multiple selections possible)}
\label{fig:vvmodels}
\end{figure}

Figure \ref{fig:vvmodels} shows the usage of different methods for verification and validation of the models. 
Testing was used most frequently (i.e., by 30\% of respondents), followed by simulation used by 24\%, manual model inspection/reviews used by 20\%, and automated model analysis used by 10\% of the respondents. Only 6\% of respondents did not perform any verification and validation of the models. 

As shown in Table~\ref{table:vvtools} only 24\% of respondents used tools to verify and validate  the models. 
These tools included Design Verifier, Model Advisor, CoCoSim, and tools listed under ``Other'' which included UPPAAL \cite{uppaal}, Jenkins \cite{jenkins}, PRISM \cite{prism}, and custom tools. 

\begin{table}[!ht]
\centering
\begin{minipage}[t]{0.48\linewidth}
\centering
\caption[Use of \VV tools for models]{Use of tools for verification and validation of the models (33 respondents)}
\label{table:vvtools}
\centering
\begin{tabular}{|p{2cm}|c|}
\hline
\multicolumn{2}{|l|}{Did you use tools for verification} \\
\multicolumn{2}{|l|}{and validation of the models?}\\
\hline
Yes & 24\% \\ %\hline
No & 55\% \\ %\hline
NA & 21\% \\ \hline
\end{tabular}%
\end{minipage}
\hfill
\begin{minipage}[t]{0.48\linewidth}
\centering
\caption{Verification and validation of autonomous functionality (32 respondents)}
\label{table:vvautonomy}
\begin{tabular}{|p{2cm}|c|}
\hline
\multicolumn{2}{|l|}{Did you specifically verify and} \\
\multicolumn{2}{|l|}{validate autonomous functionality?} \\ \hline
Yes & 56\% \\ %\hline
No & 22\% \\ %\hline
NA & 22\% \\ \hline
\end{tabular}%
\end{minipage}
\end{table}

\subsubsection{RQ5c: How were the \aucs verified and validated during development?}

\begin{figure}[!ht]
\begin{center}
\includegraphics[trim=20 225 0 200,clip,width=0.78\textwidth]{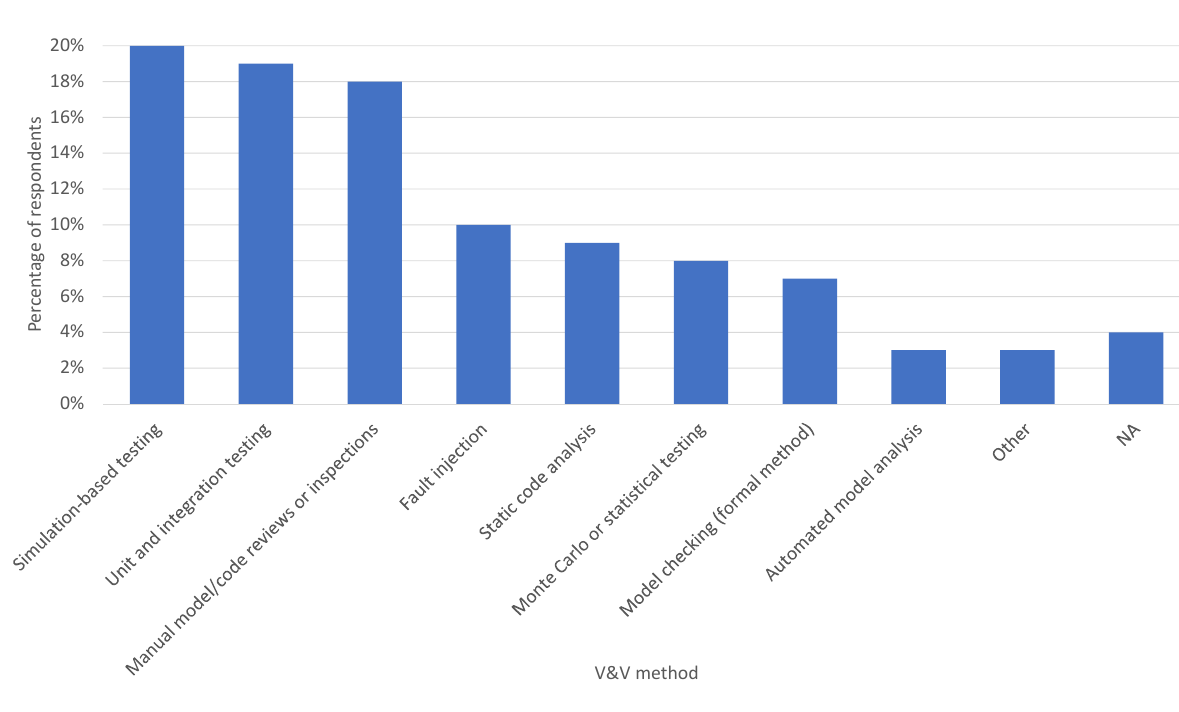}
\end{center}
\caption[Methods used for \VV of AUC during development]{Methods used for verification and validation of \aucs during development (34 respondents, multiple selections possible)}
\label{fig:vvdevelopment}
\end{figure}

As shown in Table~\ref{table:vvautonomy}, 56\% of respondents specifically verified and validated the autonomous functionality. 
The use of different methods for verification and validation of \aucs during development are summarized in Figure \ref{fig:vvdevelopment}. ``Simulation-based testing'' (20\%), ``Unit and integration testing'' (19\%), and ``Manual model/code reviews or inspections'' (18\%) were used most frequently. The least used verification and validation method was automated model analysis with only 3\%. ``Other'' methods listed by some respondents included ``Inspection of results by stakeholders'' and ``Comparing performance with human''.

\subsubsection{RQ5d: How was runtime behavior of \aucs monitored/assured?}
\begin{figure}[!ht]
\begin{center}
\includegraphics[trim=20 235 10 225,clip,width=0.78\textwidth]{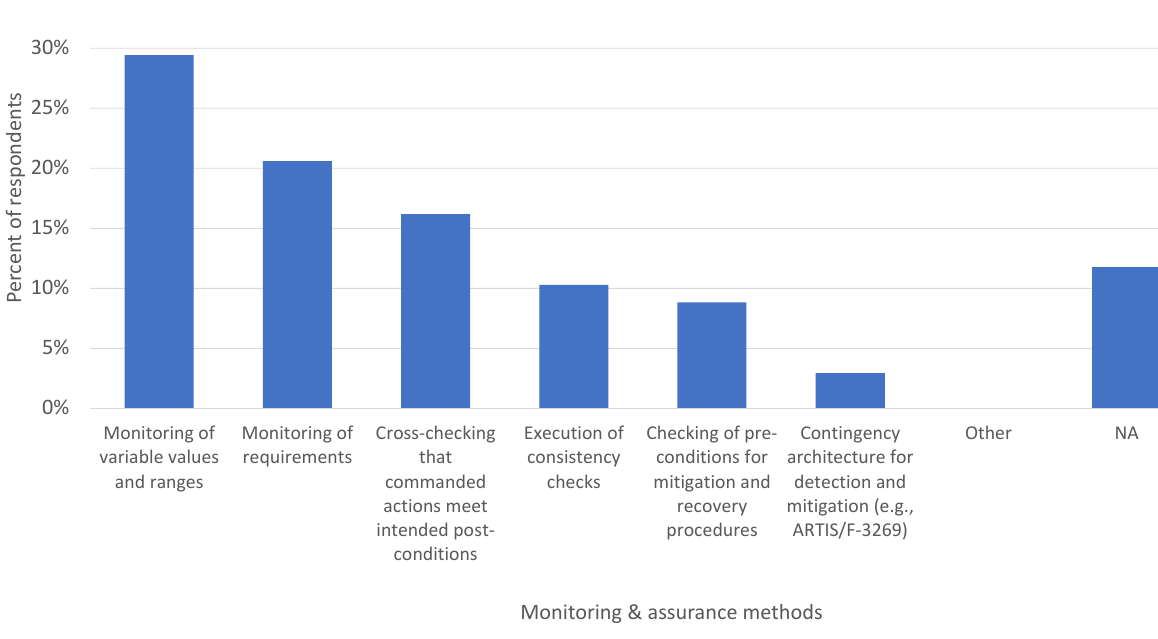}
\end{center}
\caption[How correct runtime of autonomous components were ensured]{Methods used for monitoring/assurance of \aucs runtime behavior (34 respondents, multiple selections possible)}
\label{fig:monitoraucruntime}
\end{figure}

The bar chart in Figure~\ref{fig:monitoraucruntime} presents the results related to the methods used to monitor and assure the runtime behavior of \aucs. The most frequently used methods included ``Monitoring of variable values and ranges'' with 29\%, followed by ``Monitoring of requirements'' with 21\%, and ``Cross-checking that commanded actions meet intended post-conditions'' with 16\% of the respondents. 

\begin{figure}[!ht]
\begin{center}
\includegraphics[trim= 20 235 10 210,clip,width=0.78\textwidth]{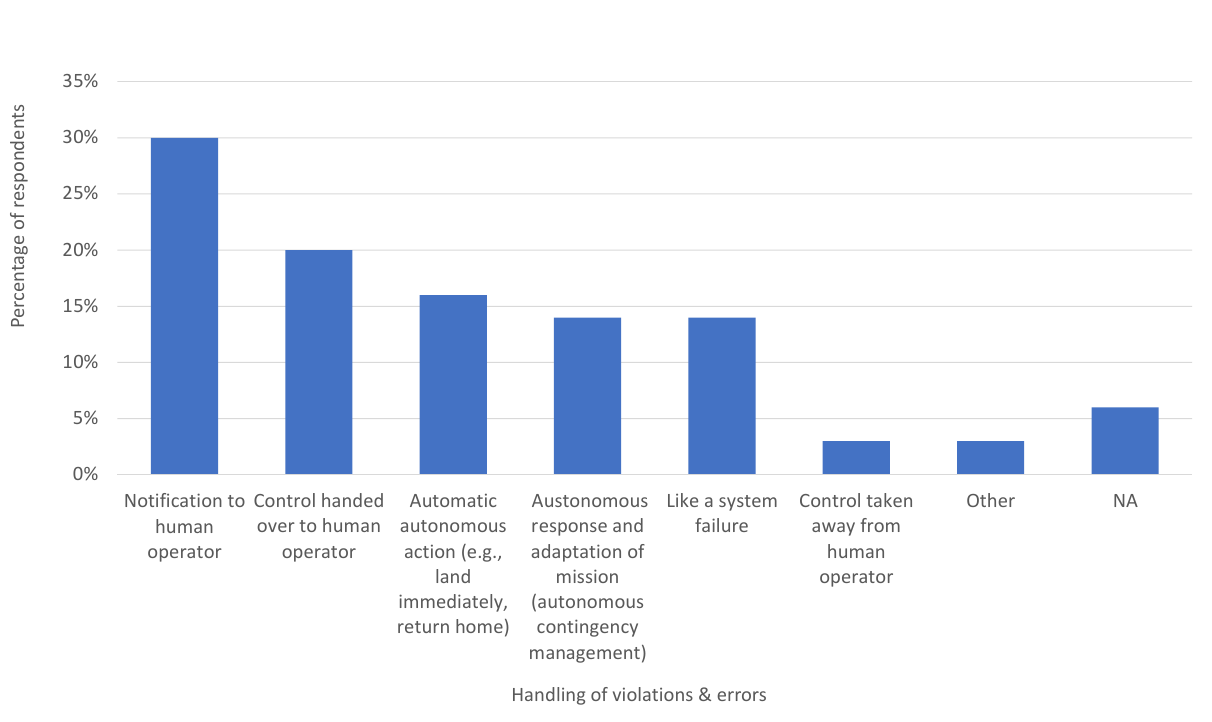}
\end{center}
\caption[Methods used for handling the violations and errors of \auc]{Methods used for handling the violations and errors of \auc (34 respondents, multiple selections possible)}
\label{fig:violationshandled}
\end{figure}

Respondents were also asked about methods used for handling the violations and errors of \aucs' behaviors. As shown in Figure~\ref{fig:violationshandled}, ``Notification to human operator'' was the most selected with 30\%, followed by ``Control handled over to human operator'', ``Automatic autonomous action'', ``Autonomous response and adaptation of mission'', and ``Like a system failure'' with 20\%, 16\%, 14\%, and 14\% of the respondents, respectively. 
Only 3\% of respondents indicated that ``Control was taken away from the human operator''. Under ``Other'', respondents listed ``Failover to layered recovery mechanisms'' and ``Probability of correctness''.

\subsubsection{RQ5e: Were the reused software artifacts verified and validated?}
The respondents were also asked if they specifically verified and validated the reused artifacts. As can be seen in Table~\ref{table:vvreuse}, only 19\% of respondents verified and validated the reused artifacts while over half (56\%) of respondents did not. 

\begin{table}[!ht]
\centering
\begin{minipage}[t]{0.48\linewidth}
%\begin{table}
\caption[\VV of reuse]{Verification and validation of reused artifacts (33 respondents)}
\label{table:vvreuse}
\centering
\begin{tabular}{|p{2cm}|c|}
\hline
\multicolumn{2}{|l|}{Did you specifically verify and} \\ 
\multicolumn{2}{|l|}{validate the reused artifacts?} \\ \hline
Yes & 19\% \\ %\hline
No & 56\% \\ %\hline
NA & 25\% \\ \hline
\end{tabular}
\end{minipage}
\hfill
\begin{minipage}[t]{0.48\linewidth}
\caption[Bugs Specific to autonomy, model-based approach, and reuse]{Bugs specific to autonomous functionality, model-based approach, and reuse (32 respondents)}
\label{table:bugs}
\centering
\begin{tabular}{|l|r|r|r|} \hline
Bugs due to                 & Yes   & No    & NA \\ \hline          
Autonomous functionality    & 28\%  & 47\%  & 25\% \\
Model-based approach        &  9\%  & 44\%  & 47\% \\ 
Reuse                       & 25\%  & 53\%  & 22\% \\ \hline
\end{tabular}%
\end{minipage}
\end{table}

%\FloatBarrier
%%%%%%%%%%%%%%%%%%%%%%%%%%%%%%%%%%%%%%%%%%%%%%%%%%%%%%%%%%%%%%%%%%%%%%%%%%%%%%%%%%%%%%
\subsection{Bugs}
%%%%%%%%%%%%%%%%%%%%%%%%%%%%%%%%%%%%%%%%%%%%%%%%%%%%%%%%%%%%%%%%%%%%%%%%%%%%%%%%%%%%%%
\subsubsection{RQ6a: Where there any bugs specific to autonomous functionality, model-based approach, and/or reuse?}

Our survey also explored the existence of bugs specific to autonomous functionality, model-based approach, and reuse of software artifacts. The responses are summarized in Table~\ref{table:bugs}. 
28\% of respondents reported finding bugs specific to autonomous functionality. 
As expected, the ratio of specific bugs found vs.\ no-specific bugs found was high (approximately 1:2) for autonomous functionality. 
High complexity of the autonomous functionality code and novel/ unusual algorithms may be the likely reasons.
Several respondents provided reasons for bugs related to autonomous functionality, which included overfitting/ underfitting and failures of commercial-off-the-shelf components. 

Only 9\% of respondents reported finding bugs specific to model-based approach. The ratio of specific bugs found vs.\ no-specific bugs found was relatively low (roughly 1:5), as expected. 

Surprisingly, the percentage of reuse specific bugs was high (i.e., 25\%), as well as the ratio of specific bugs found vs.\ no-specific bugs found (approximately 1:2). This means that a substantial number of reuse-specific bugs existed and also seemed to occur, in contrast to the typical assumption that reuse would substantially lower the number of bugs.
Some respondents listed specific reasons for bugs attributed to reuse, which included version conflicts, incompatibilities, adaptation of existing software produced unintended consequences, and bugs that were not corrected in the previous release.

%%%%%%%%%%%%%%%%%%%%%%%%%%%%%%%%%%%%%%%%%%%%%%%%%%%%%%%%%%%%%%%%%%%%%%
\section{Threats to Validity}
\label{ch:ThreatsToValidity}
%%%%%%%%%%%%%%%%%%%%%%%%%%%%%%%%%%%%%%%%%%%%%%%%%%%%%%%%%%%%%%%%%%%%%%

In this section, we describe the threats to the validity of this study and the measures taken to
mitigate them.

\textbf{Construct validity} addresses whether we are testing what we intended to test. 
One threat to construct validity is the use of inconsistent and imprecise terminology. To avoid this threat, in the survey and the paper, we provided the definitions of the terms being used. 
To ensure if we test what we intended to test, we carefully designed the survey questionnaire and used a pilot study to verify and validate it and to subsequently augment and improve it. 
The survey relied on voluntary participation, which introduces the risk of self-selection bias. We tried to address this threat to validity by reaching to experts from different domains, both from industry and academia, worldwide using different means of communication.  
Humans are know to have evaluation apprehension (i.e., a tendency to try to look better or fear of being evaluated). For example, participants may deliberately exaggerate or downplay their skills, affecting the validity of the survey results. Furthermore, the fear of negative consequences may result in a reluctance to report their true experiences accurately. Many of these social threats to construct validity were mitigated by keeping the survey anonymous, and allowing the respondents to skip any question.

\textbf{Internal validity} concerns influences that can affect the variables and metrics without researchers’ knowledge. 
The survey required participants to recall and self-report their experiences related to development of autonomous systems. Participants' recollections may be influenced by memory limitations or biases, potentially affecting the accuracy of their responses. 
Such biases affect individual participants who responded to the survey. In our survey with more than 100 respondents, we believe that such biases canceled themselves out.

\textbf{Conclusion validity} concerns the ability to draw correct conclusions. 
One of the threats to conclusion validity is the sample size. We distributed the survey widely and collected responses from 110 experts who have used autonomous systems and/or MBSwE in their projects. The answers related to the autonomous functionality were collected for 58 AUCs. In general, these are fairly large sample sizes that allow drawing valid conclusions. Some survey questions, however, have smaller sample sizes because answering each question was not mandatory. For clarity, the results are annotated with the number of participants who answered that specific question.   

\textbf{External validity} concerns the generalizability of results. 
The results presented in this paper are based on opinions of experts on autonomous systems from different domains and world regions, both from industry and academia which ensures representative population and some generalizability of the results. Nevertheless, we cannot claim that the results based on one survey would be valid for all autonomous systems. 

%%%%%%%%%%%%%%%%%%%%%%%%%%%%%%%%%%%%%%%%%%%%%%%%%%%%%%%%%%%%%%%%%%%%%%
\section{Recommendations and Conclusion}
\label{ch:summary}
%%%%%%%%%%%%%%%%%%%%%%%%%%%%%%%%%%%%%%%%%%%%%%%%%%%%%%%%%%%%%%%%%%%%%%

This paper presents the first part of a longitudinal study of software systems that have autonomous functionality and may employ MBSwE and reuse.  
The empirical results are based on data collected using an online survey which was conducted in 2019. The respondents to the survey were from different industry domains, worldwide. 
As main contributions of this paper, we (C1) assessed the state-of-the-practice of developing autonomous systems at the time of the survey, (C2) identified and quantified the benefits and challenges of autonomy and reuse, (C3) explored the processes and standards used to develop autonomous systems, and (C4) investigated the verification and validation of the autonomy, models, and reuse.

Our survey revealed that most project with autonomous functionality employed high levels of autonomy (see Figure~\ref{fig:auctaskautonomy}).
Based on the findings presented in this paper, it can be concluded that the {\em safety-critical systems with a substantial degree of autonomy\/} have been successfully developed in the industry and that implementation of tasks with full autonomy was within the realm of the existing software technology. 
%Challanges
%
However, development of autonomous functionality was not without challenges. Among the respondents of our survey, the most challenging was to deal with the {\em system complexity\/}, followed by the high level of {\em environment uncertainty\/} and the difficulty achieving the {\em desired level of autonomy}. 
Surprisingly, {at the time the survey was conducted (i.e., 2019)}, the development of autonomous functionality was still dominated by {\em traditional algorithms\/} (i.e., rule-based algorithm, planning systems/languages, and statistical and filtering methods), with only 12\% of AUCs using machine learning approaches. 
More detailed summaries of our findings, grouped by research question, are given in Table~\ref{table:findings}. 

\begin{table}[!ht]
\centering
\caption{Summary of the main findings for the research questions given in Table~\ref{table:research_questions}}
\label{table:findings}
\begin{tabular}{|p{1cm}|p{13cm}|}
\hline
\multicolumn{2}{|c|}{\cellcolor{lightgray}\textbf{Findings 1: Where? What? How? and Who?}}                                                                   \\ \hline
RQ1a & Where: Space industry dominated with 40\%, followed by aviation with 16\%, military with 13\% and automotive with 9\% of respondents.  \\ \hline
RQ1b & What: 45\% of respondents indicated catastrophic or hazardous level of safety criticality, followed by 39\% with major safety criticality of their systems. \\ \hline
RQ1c & What: C and C++ or a mixture thereof dominated as implementation languages, both during development and deployment, followed by use of Python. 
MATLAB was also used, but much more frequently during development than during deployment. \\ \hline
RQ1d & How: Both during development and deployment, over one third of the code was developed from scratch (i.e., 36\% and 38\%, respectively). Next most common were using customized existing code libraries (with 23\% both during development and deployment) and using existing code libraries (with 20\% and 17\%, respectively for development and deployment). \\ \hline
RQ1e & How: Special hardware and cloud services were not used frequently. Thus, only 16\% during development and 19\% during operation used special hardware, and only 15\% used cloud services.\\ \hline
RQ1f & Who: The two major groups of respondents' roles included ``Design'' and ``Research'', followed by ``Model development'', ``Programming'', ``Software and system integration'', ``Testing/QA/V\&V'', and ``Project management''.                                             \\ \hline %\hline
\end{tabular}
\end{table}

\newpage

\begin{table}[ht]
\centering
\caption*{Summary of the main findings for the research questions given in Table~\ref{table:research_questions} (continuation)}
\begin{tabular}{|p{1cm}|p{13cm}|}
\hline
\multicolumn{2}{|c|}{\cellcolor{lightgray}\textbf{Findings 2: Details on Autonomy}} \\ \hline
RQ2a & 38\% of \aucs were developed using \MBSE.
\\ \hline
RQ2b & For each task (i.e., Information acquisition, Information analysis, Decision and action selection, and Action implementation), a significant percentage of \aucs (i.e., from 50\% to 67\%) operated with the computer having a full autonomy. \\ \hline
RQ2c & The majority of \aucs used traditional algorithms and methods like rule-based algorithms (35\%), planning systems/ languages (22\%), and statistical and filtering methods (18\%). Surprisingly, Machine learning approaches were only used for 12\% of \aucs. 
\\ \hline
RQ2d & For 53\% of \aucs the requirements were specified in the same way as for non-\aucs. Natural Language was used more than twice as often as formal specifications. The majority of \aucs (i.e., 63\%) were developed using requirements with the same level of details as the non-\aucs.  \\ \hline
RQ2e & When major and moderate challenges are considered together, ``System complexity'' was the most challenging (for 67\% of the \aucs), followed by  ``High level of environment uncertainty'' and ``Achieving the desired level of autonomy'' (for 57\% and 40\% of the \aucs, respectively).  \\ \hline
\multicolumn{2}{|c|}{\cellcolor{lightgray}\textbf{Findings 3: Details on Reuse of Software Artifacts}}  \\ \hline
RQ3a & For both \aucs and non-\aucs, software code was the most reused artifact, with over 60\% of the responses indicating 30\% or more of the code being reused.   \\ \hline
RQ3b & When major and moderate negative aspects of reuse were considered together,``Hindering new ideas’’ and ``Added complexity because of reuse’’ were most significant with 25\%, followed by ``Additional cost due to reuse sustainability’’ with 21\%. 
\\ \hline
RQ3c & When major and moderate difficulties due to reuse were considered together, 
``Uncertain operational conditions / environment’’ led to most difficulties (40\%), followed by 
``Verifying and validating reused software'' (34\%),
``Lack of planning for reuse in advance’’ (32\%) and 
``Integrating the reused parts into the development environment'' (31\%). \\ \hline
RQ3d & 63\% of respondents reported increase productivity, 40\% reported increased quality, and only 37\% reported decreased cost due to reuse.  \\ \hline \hline
\multicolumn{2}{|c|}{\cellcolor{lightgray}\textbf{Findings 4: Processes and Standards}}    \\ \hline
RQ4a & Twice as many respondents used Agile  than Waterfall life-cycle process (i.e., 38\% vs.\ 19\%).  \\ \hline
RQ4b & Almost half of the respondents (i.e., 48\%) did not use any modeling standard and about a quarter of the respondents (i.e., 26\%) did not follow any coding standard.  \\ \hline
RQ4c & Only 24\% of systems went through certification and in only 26\% of cases \aucs were part of a certified system.
\\ \hline \hline
\multicolumn{2}{|c|}{\cellcolor{lightgray}\textbf{Findings 5: Verification \& Validation}}    \\ \hline
RQ5a & Most of the respondents verified and validated the performance (83\%) and the correctness (78\%) of the software. Robustness and safety were verified and validated in 61\% and 56\% of the cases, respectively. Only 11\% verified and validated security.    \\ \hline
RQ5b & Testing was used most frequently for verification and validation of the models (i.e., by 30\% of respondents), followed by simulation with 24\%, manual model inspection/reviews with 20\%, and automated model analysis with 10\%. Only 6\% of respondents did verify and validate the models.  \\ \hline
RQ5c & During development, \aucs were most frequently verified and validated using simulation-based testing (20\%), unit and integration testing (19\%), and manual model/code reviews or inspections (18\%). Automated model analysis was used only by 3\% of the respondents. \\ \hline
RQ5d & Most frequently used methods for monitoring / assuring \aucs' runtime behavior included ``Monitoring of variable values and ranges'' (29\%), ``Monitoring of requirements'' (21\%) and ``Cross-checking that commanded actions meet intended post-conditions'' (16\%). \\ \hline
RQ5e & Only 19\% of respondents verified and validated the reused artifacts. \\ \hline \hline
\multicolumn{2}{|c|}{\cellcolor{lightgray}\textbf{Findings 6: Bugs}}    \\ \hline
RQ6a & 28\% of respondents reported finding bugs specific to autonomous functionality. Only 9\% of respondents reported finding bugs specific to model-based approach. Surprisingly, the percentage of reuse specific bugs was high (i.e., 25\%). \\ \hline
\end{tabular}
\end{table}

\newpage
Based upon our findings, the following recommendations 
can support and help to further enhance the development of autonomous systems: 
\begin{itemize}

\item {\bf High complexity of AUCs and environmental uncertainties require careful upfront study and planning}.
This has been a major issue according to our findings.
Because there are often big and unclear expectations about system capabilities and
performance, detailed requirements and operational envelopes should be 
defined early in the process. 

\item {\bf Model-based Software Engineering can help toward successful 
development of an AUC}. As 37\% of respondents used MBSwE, this is a strong
indication of the usefulness of \M. 
The decision to use \M should be made early in the process and suitable
tools set up and made available to the project team.

\item {\bf Projects that incorporate autonomy should be encouraged 
to use modeling and coding standards.}
Our survey showed that modeling standards were used by roughly half of the projects whereas coding standards were applied in 74\% of the projects.
Enforcements of modeling and coding standards can contribute to project's success and may also help with reuse of software artifacts.

\item {\bf Certification of AUCs is still a tough issue.}
Although many of the projects in our survey contained safety-relevant AUCs,
only about a quarter of the AUCs went through certification.
This might have been due to unavailability of standards tailored toward AUC certification at the time our survey was conducted. 
Raising awareness of new and upcoming certification standards for
autonomous systems and AUCs (e.g., IEEE P7009 for failsafe design \cite{IEEE7009_2024})
and for learning-enabled systems (e.g., EASA Concept Paper 
\cite{EASAL1concept2021,EASAL1-2concept2024}) and
aligning in-house development and quality assurance processes with such guidelines
can streamline future projects that require certification.

\item {\bf Reuse of software artifacts for development of autonomous systems can be helpful, but requires careful planning and considerations.}
Over 60\% of respondents to our survey reported increased productivity 
due to reuse.
However, only 40\% or less experienced positive impacts of reuse 
on quality and cost. 
Note that reuse always requires up-front investments to prepare artifacts
for reuse.

\item {\bf Verification and validation should be performed on reused software artifacts as well} since surprisingly high percentage of respondents (i.e., 25\%) experienced reuse specific bugs. Careful preparation of artifacts (e.g., documentation, attached certification, component-specific test cases) for later reuse is therefore important and can be seen as a good investment.

\item {\bf Software verification and validation should go beyond performance and correctness; it should also include robustness, safety, and security.} 
Our survey results showed that performance and correctness were verified most frequently (with 83\% and 78\%, respectively), followed by robustness and safety (with 61\% and 56\%, respectively). Verification and validation of security was rare, done by only 4\% of the respondents. 

\item {\bf The collection of empirical knowledge about software bugs specific to autonomy would be very valuable and has a potential to cost effectively increase the quality of AUCs.}
Based on the survey responses, software contained bugs specific to autonomy more often than bugs specific to model-based approaches (i.e., 28\% versus 9\% of responses). 
For details see e.g., \cite{goseva2023COMPSAC}.
\end{itemize}

Our current work is focused on conducting the second part of our longitudinal study whose goals are to explore 
how the state-of-the-practice has evolved over time and 
if the challenges and level of machine learning usage in autonomous systems remain the same or may have changed over the last six years.

\section*{Acknowledgments} 
This work was funded by the NASA Software Assurance Research Program (SARP) in fiscal years 2019 and 2020. The authors would like to thank the colleagues from the NASA Ames Research Center and other colleagues who provided their feedback on the survey questionnaire. The authors would also like to thank all respondents to the survey for sharing their information, experience, and opinions.

\bibliographystyle{IEEEtran}

\end{document}